\documentclass[11pt]{article}

\usepackage{float}
\usepackage{geometry}
\newcommand{\fref}[1]{{Fig.~\ref{fig:#1}}}
\newcommand{\Fref}[1]{{Figure~\ref{fig:#1}}}

\usepackage{amsmath}
\usepackage{xcolor, overpic, rotating}

\usepackage{gensymb}

\title{\LARGE{\vspace{-.25in}\textbf{Near Wake Dynamics of a Cross-Flow Turbine Array}}}
\author{Isabel Scherl$^{1*}$, Abigale Snortland$^1$, Steven L. Brunton$^1$, and Brian Polagye$^1$\\
\footnotesize{$^1$ Department of Mechanical Engineering, University of Washington, Seattle, WA 98195, USA}\\}

\begin{document}

\maketitle

\begin{abstract}
Cross-flow turbines, also known as vertical-axis turbines, convert the kinetic energy in moving fluid to mechanical energy using blades that rotate about an axis perpendicular to the incoming flow. In these experiments, the performance and wake of a two-turbine array in a fence configuration (side-by-side) was characterized. The turbines were operated under coordinated control, a strategy characterized by synchronous rotation rates with a mean phase difference. Measurements were made with turbines co-rotating, counter-rotating with the blades advancing upstream at the array midline, and counter-rotating with the blades retreating downstream at the array midline. From the performance data, we find individual turbine and array efficiency depend significantly on rotation direction and phase difference. Variations are also observed in the wake flow field and, using these data, we hypothesize how rotation direction and phase influence interactions between adjacent turbines.

\end{abstract}

\section{Introduction}

Cross-flow turbines harness sustainable energy from wind and marine currents~\cite{salter2012coe,eriksson2008jrse} for a range of applications and scales. Often referred to as vertical-axis turbines in the wind energy space, cross-flow turbines rotate about an axis perpendicular to the incoming flow.

While the performance of a cross-flow turbine depends on its geometry (e.g., number of blades~\cite{castelli_2012,hunt2023parametric}, chord-to-radius ratio~\cite{parker2016eif}, blade profile~\cite{carrigan_2012}, and preset pitch angle~\cite{chen_2013}) and operating condition (e.g., channel blockage~\cite{ross2020re}, Reynolds number~\cite{bachant2016energies}), all fixed-pitch cross-flow turbines experience similar unsteady fluid-structure interaction. Specifically, as a cross-flow turbine blade rotates, the angle of attack of its blades is constantly changing, resulting in unsteady boundary layer separation and dynamic stall, an active area of research for which a universal model does not yet exist \cite{corke2015ar,le2022dynamic, dave2023analysis}. 
Consequently, they exhibit high-dimensional, complex dynamics ~\cite{bachant2016plos, posa2016johff,snortland2023cycle} that have been utilized to develop advanced control strategies with non-constant rotation rates to increase the power produced by an individual rotor~\cite{strom2017natenergy, dave2021aiaa, athair2023intracycle}. 
These unsteady dynamics are not present in traditional axial-flow turbines, though similar dynamics can be seen in energy harvesters that utilize heaving and pitching foils~\cite{ribeiro2021prf}. 
Following stall separation, periodic structures are advected into the wake~\cite{dunne2015eif, dave2021jrse}. 
Cross-flow turbine wake dynamics are broadly characterized as separated flows with significant three dimensionality~\cite{ryan2016eif, posa2022re, bachant2015jot}. However, previous work has demonstrated that useful insights about near-wake dynamics can be obtained from two-dimensional planes normal to the blade~\cite{simao2009visualization, tescione2014re}.

Large-scale energy harvesting by cross-flow turbines typically requires arrays of rotors arranged in rows. A significant advantage of cross-flow turbines is that they can interact beneficially, producing more power together than in isolation, outperforming an equivalent axial-flow turbine array~\cite{whittlesey2010bioinspiration, kinzel2012jot}. 
In fact, larger arrays are estimated to have a power density per unit land area 6-9 times greater than axial-flow turbine arrays~\cite{dabiri2011jrse}. 

When expanding from a single turbine to a multi-turbine array, the system becomes increasingly high dimensional, turbulent, and nonlinear. The flow induced by each rotor depends heavily on the relative positions of the turbines in the array, rotation rates relative to the incoming flow, the direction of rotation for each turbine, and any mean phase difference between the turbines~\cite{scherl2020jrse}.
These factors increase the dimensionality of the problem relative to an individual turbine due to the potential for multiple interaction mechanisms between turbines in the array. For example, mutual interactions can occur between turbines in relatively close proximity, either through mean flow alteration or coherent structure interaction, as well as downstream wake interaction at greater separation distances after coherent structures have dissipated.

Dense array interactions between nearby turbines have analogues to flow control in other fluid systems, such as fish schooling ~\cite{fish2006ar, wu2011arfm, verma2018pnas}, heaving and pitching foils for energy generation~\cite{miller2016dfd, ribeiro2021prf}, and bird flight in both the clap and fling mechanism~\cite{ellington1984aerodynamics,weis1973quick} and flocking~\cite{portugal2014nature, wu2011arfm}. 
Prior investigations of cross-flow turbine arrays in field experiments~\cite{dabiri2011jrse,kinzel2012jot,kinzel_2015}, laboratory experiments~\cite{brownstein2016jrse, brownstein2019energies, ahmadi_2016, vergaerde2020re, doan2020jmse, scherl2020jrse,scherlparameter,jodai2023wind, huang2023experimental}, and simulation~\cite{duraisamy_2014,zanforlin2016re,bremseth_2016,durrani_2011,chen_2017, de2018towards, sahebzadeh2020towards, gauvin2022} have demonstrated that power output increases when the rotors are arranged in a side-by-side configuration. Further, as the rotor spacing in side-by-side rotors decreases, array efficiency increases~\cite{zanforlin2016re, jin2020oe, scherl2022optimization}.

At greater distances downstream of the rotor, wake evolution and interaction with subsequent rows is also of considerable interest, but wake studies of multiple cross-flow turbines are less common than for individual turbines. 
Zanforlin et al.~\cite{zanforlin2016re} simulated pairs of counter-rotating turbines with varied rotor spacing, non-dimensional rotation rate (tip-speed ratio), and wind direction. They note that turbines placed side-by-side induced beneficial mean flow alterations in the cross-stream direction, increasing lift on the blade and turbine torque. In the wake, they found that the individual deficit regions contract downstream of the rotors. 
In Gauvin-Tremblay et al.~\cite{gauvin2022hydrokinetic}, the authors utilize a simplified computational turbine model to simulate pairs of rotors. Their primary finding is that staggering rotors does not enhance performance and that side-by-side configurations are preferred. 
Posa et al.~\cite{posa2019ijhf} simulated flow fields for co-rotating and counter-rotating pairs of turbines. 
This work was then extended to include blade-tip effects and wake recovery, finding that three-dimensional edge effects accelerate wake recovery~\cite{posa2022re}. 
In both of these computational studies, the turbines are operated in a synchronous manner without any phase offset (i.e., blades on both turbines are always at identical azimuthal positions).  
On the other hand, Jin et al.~\cite{jin2020oe} perform a computational study on rotation scheme and rotor phase difference. They found that the main driver of wake dynamics is turbine rotation scheme and that effects from phase difference are secondary.  
Performance and wakes of turbine pairs offset in the streamwise direction have also been explored. Brownstein et al. ~\cite{brownstein2016jrse} found offset arrays to have increased efficiency relative to the individual turbines, and the time-averaged flow fields showed varying dynamics dependent on rotation scheme. 
Finally, Lam et al.~\cite{lam2017energy} experimentally measured asymmetric wake skew patterns in paired turbines that are consistent with those seen in individual rotor wakes.

Consequently, our understanding of rotor interactions and array dynamics, such as the the effect of turbine phase and rotor rotation schemes on wake dynamics and performance, is growing, but limited. Additionally, there has been little exploration of phase-resolved wake dynamics for arrays, which is expected to be considerable, given the strong phase-dependence seen in studies of individual turbines ~\cite{strom2022jfm}. 
On the whole, we are limited by both the measurable quantities in our experiments and the inherent difficulty in disambiguating flow structures and their effects on turbine dynamics and control. 
In other words, determining causality between performance and flow field alterations, not just correlation, remains a central challenge in understanding both individual turbines and arrays through experiments.

In this work, we perform an experimental evaluation of the performance and wake of a two-turbine array under a ``coordinated control'' strategy with controlled mean phase offsets and constant rotation rate. This comprehensively addresses the potential operational modes and correlations between power production and wake evolution. 
Specifically, turbine pairs are operated co-rotating, counter-rotating with the blades advancing (or traveling upstream) as they approach the mid-line of the array, and counter-rotating with the blades retreating (or traveling downstream) as they approach the mid-line of the array. The array midline is the streamwise line at the middle of the array (i.e., the closest point of approach between the two rotors).
First we present the array performance for each rotation scheme over the full range of phase differences. Then we explore mean wake dynamics across rotation schemes for selected phase differences. Lastly, we compare phase resolved wake dynamics of the counter-rotating arrays. 

\section{Methods}

This section outlines the techniques used in this work, including information on turbine dynamics and experimental methods.
Within the turbine dynamics, we outline the array configurations and phase offset definitions. The experimental methods describe both performance and particle image velocimetry (PIV) measurements and the environmental conditions of the experiments.

\begin{figure}

\begin{overpic}[width=0.99\textwidth]{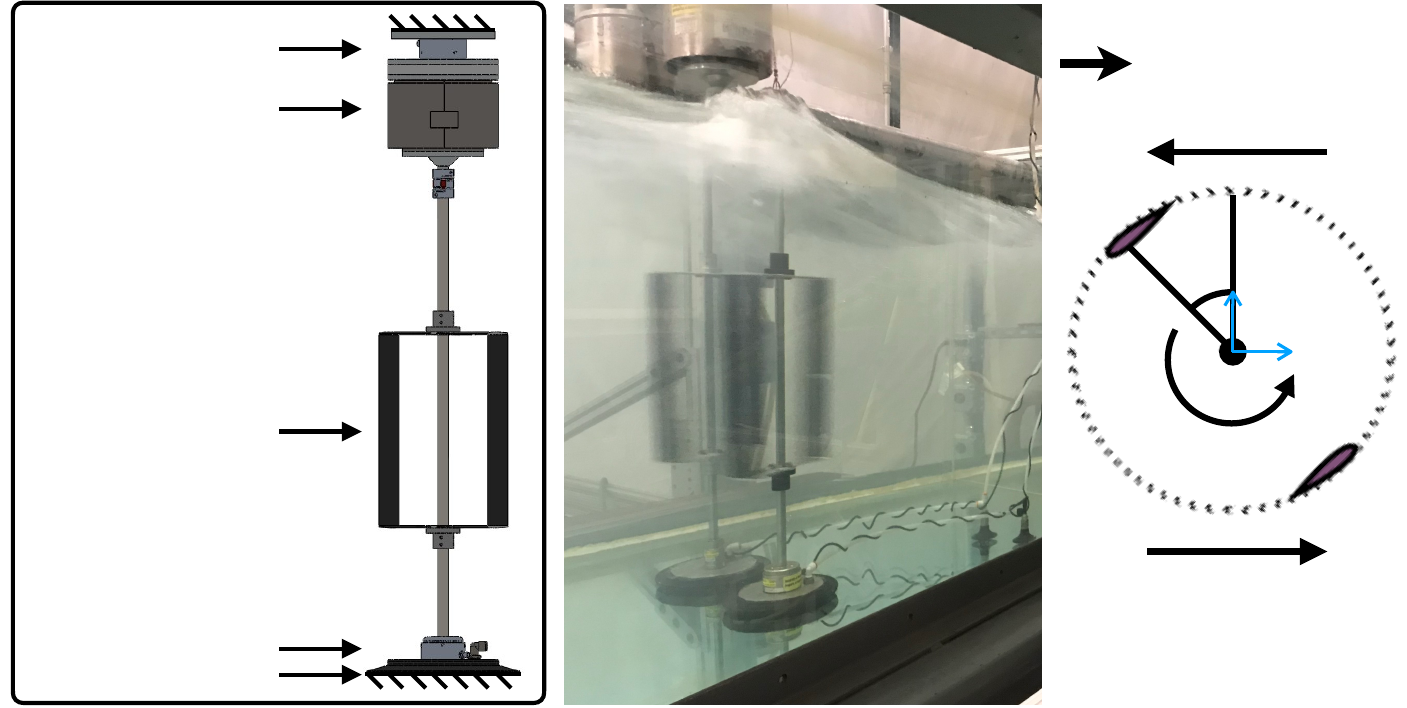}
\put(91.5,26){$\scriptstyle{F_x}$}
\put(88,30){$\scriptstyle{F_y}$}
\put(92,21){$\scriptstyle{\tau}$}
\put(8,46.5){\textbf{load cell}}
\put(10.25,42.25){\textbf{motor}} 
\put(2,19.25){\textbf{turbine rotor}}
\put(8,4){\textbf{load cell}}
\put(1.75,2){\textbf{vacuum plate}}

\put(81,43.5){\textbf{Advancing}}
\put(81,7){\textbf{Retreating}}

\put(85,30){$\theta$}

\put(0,51.5){\small{\textbf{(a)}}}
\put(39.5,51.5){\small{\textbf{(b)}}}
\put(75,51.5){\small{\textbf{(c)}}}

\put(75,48){$U_\infty$}
\end{overpic}
\vspace{-.15in}
\caption{(a) Turbine assembly with load cells, motor, rotor, and vacuum plate labeled (b) photo of turbine assemblies operating in the flume (c) diagram of turbine showing the advancing and retreating portions of the rotation}\label{fig:exp}
\end{figure}
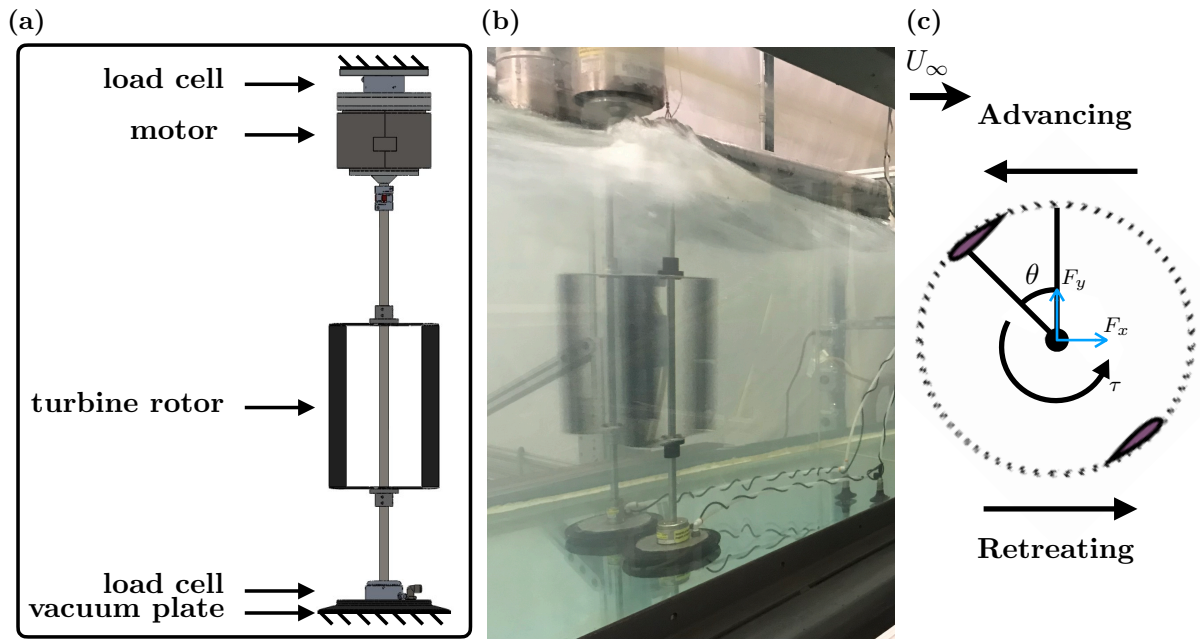

\subsection{Cross-flow turbine performance and dynamics}

To evaluate the performance of a single turbine, we use the coefficient of performance $(C_P)$ which is the ratio of the power produced by the turbine to the kinetic power in the free-stream flow passing through the rotor's projected area. The coefficient of performance is expressed as 
\begin{equation}
C_p = \frac{P}{\frac{1}{2}\rho U_{\infty}^3HD} = \frac{\omega \tau}{\frac{1}{2}\rho U_{\infty}^3HD},
\end{equation}
where $P$ is turbine's mechanical power, $\rho$ is the fluid density, $U_\infty$ is the freestream flow velocity, $H$ is the turbine height, $D$ is the turbine diameter, $\omega$ is the turbine rotation rate, and $\tau$ is the turbine torque. {A similar non-dimensionalization is performed for the lateral force and thrust forces
\begin{equation}
C_{F,x} = \frac{F_x}{\frac{1}{2}\rho U_{\infty}^2HD}
\end{equation}

\begin{equation}
C_{F,y} = \frac{F_y}{\frac{1}{2}\rho U_{\infty}^2HD}
\end{equation}
where $F_x$ is the force on the rotor in the streamwise direction (thrust force) and $F_y$ is the force in the cross-stream direction (lateral force). 

Rotation rate is non-dimensionalized as the tip-speed ratio: 
\begin{equation}
\lambda =\frac{\omega R}{U_{\infty}}, 
\end{equation}
where $R$ is the turbine radius. The $C_P-\lambda$ curves have a global maximum (i.e., peak efficiency) at an optimal $\lambda$ set point (\Fref{cp_phi}(a)). 

As previously discussed, cross-flow turbines have high-dimensional, periodic dynamics~\cite{parker2016eif, strom2017natenergy}. As a turbine rotates, the angles of attack of the blades are constantly changing and structures are periodically shed from the rotor. A single rotation can be bisected into the advancing portion, where the blade is traveling upstream ($\theta = 270\degree - 90\degree$), and retreating portion, where the blade is traveling downstream ($\theta = 90\degree - 270\degree$). The advancing and retreating sections of rotation are labeled in \Fref{exp}(c). 

We examine dimensionless power and force as phase-averages at each azimuthal position ($\theta$), as well as time-averaged values where the measurements are averaged over all azimuthal positions and rotations at a given operational scheme. Moving from the reference frame of a single rotor to the array, power is defined as a positive quantity when hydrodynamic torque acts in the direction of rotation and thrust forces always act in the downstream direction (\Fref{exp}c), so the sign convention for these terms is equivalent between reference frames. However, to avoid the sign reversal for counter-rotating turbines in a global reference frame overshadowing inter-rotor differences, lateral forces are always presented in the reference frame of an individual rotor. Consequently, when considering the array performance in aggregate, we utilize the array mean of individual rotors. 

At an array level, we generally consider dimensionless power and force as a function of phase offset ($\phi$). Array performance is described by a time average at a given $\phi$ (i.e., $<C_P>$) and we quantify uncertainty and variability through the standard deviation of individual cycle averages. To highlight the performance differences as a function of phase offset, rather than control strategy, we use a mean subtracted form:

\begin{equation}
\Delta C_{P} = C_{P} - <C_{P}>.
\end{equation}

\noindent Here, $<C_{P}>$ is the array-average performance over all phase offsets for a given control strategy (i.e., rotation direction). This mean subtracted form is applied to individual turbines, as well as the array, and is also applicable to the force coefficients.}

\begin{figure}

\begin{overpic}[width=0.99\textwidth]{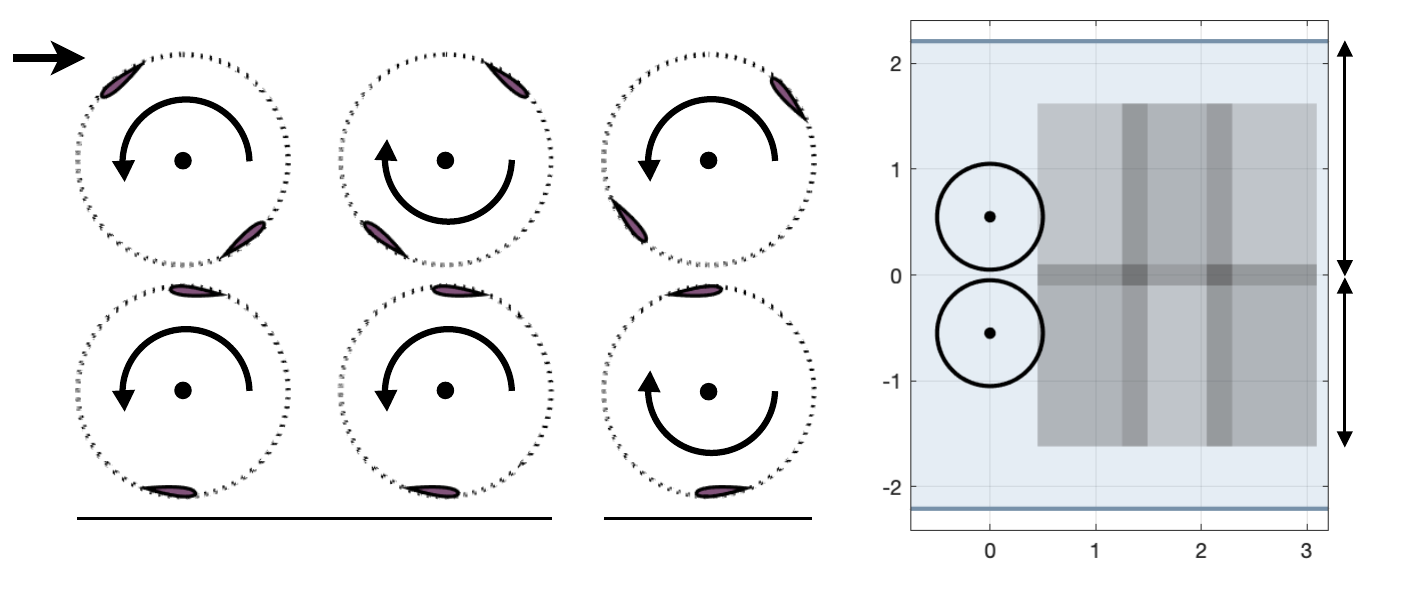}
\put(80.5,0){$\scriptstyle{X/D}$}
\put(97,16){$\scriptstyle{1.6D}$}
\put(97,30){$\scriptstyle{2.2D}$}

\put(0,16){T1}
\put(0,30){T2}

\put(61,20.75){\begin{sideways}$\scriptstyle{Y/D}$\end{sideways}}

\put(0,39.5){$U_{\infty}$}

\put(11.5,40){Co-}
\put(25.5,40){Advancing}
\put(46,40){Retreating}

\put(18,1){$\phi = \theta_2 - \theta_1$}

\put(43,1){$\phi = \theta_2 - \theta_1 - 90\degree$}

\put(0,43.5){\small{\textbf{(a)}}}
\put(62,43.5){\small{\textbf{(b)}}}

\end{overpic}
\vspace{-.15in}
\caption{(a) Rotation schemes, ``co-''rotating, ``advancing'' counter-rotating, and ``retreating'' counter-rotating shown with their respective phase difference calculations. Turbine 1 (``T1'') is located at a $Y/D$ of -0.55  and Turbine 2 (``T2'') is located at a $Y/D$ of 0.55. The array midline is located at $Y/D = 0$. For all cases, Turbine 1 is at $\theta_1 = 0\degree$ and the phase difference shown in the schematic is $\phi = 45\degree$, (b) turbine positions and PIV fields shown in the flume coordinate frame normalized by turbine diameter. }\label{fig:exp1}
\end{figure}
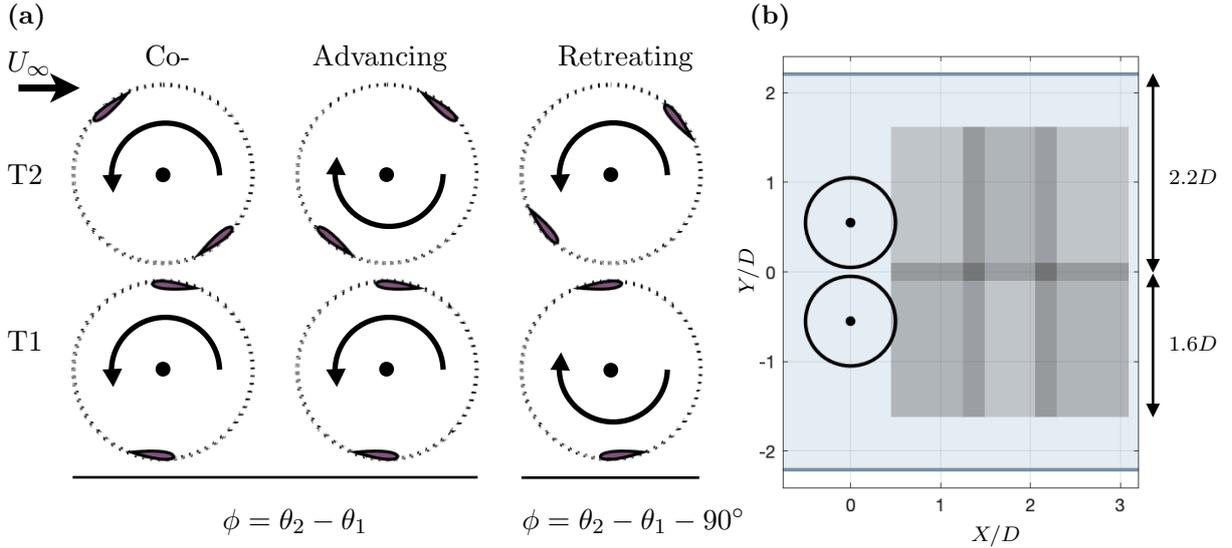

\subsection{Array Configurations, Control, \& Symmetry}

The turbines were tested co-rotating (rotating in the same direction) and counter-rotating (rotating in opposite directions). 
Within a single rotation of a cross-flow turbine there is inherent asymmetry. For half of the cycle, the blade is advancing and for the other half the blade is retreating (\Fref{exp}c). 
Given the bi-directional travel of a blade over the course of a single rotation, there are two possible counter-rotating fence (side-by-side) array configurations. 
These cases are defined by which direction the blades are traveling at the array midline (i.e., their closest point of approach or $Y/D$ = 0). 
For the retreating counter-rotating case, the blades are traveling in the same direction as the freestream flow or retreating when they are approaching the array mid-line ($Y/D = 0$). Conversely, for the advancing counter-rotating case, the blades are traveling against the direction of the freestream flow or advancing as they approach the mid-line of the array. 
The three rotation schemes are shown in \Fref{exp1}(a). The phase difference or the average difference in phase between the two turbines is explicitly defined in this figure and all arrays are shown with Turbine 1 at a phase of $\theta_1 = 0\degree$ and a phase difference of $\phi = 45\degree$. 
For each of the rotation schemes, performance was characterized across phase differences from $\phi = 0\degree-180\degree$ in increments of $6\degree$, reducing the experimental space due the symmetry of a two-bladed rotor.

\subsection{Turbine Performance Measurements}

The data presented in this work was collected in the Alice C. Tyler flume at the University of Washington. This recirculating water channel has a width of 76 cm. The test conditions consisted of a water depth of 52.5 cm, temperature of  23.5\degree C, and freestream flow velocity ($U_\infty$) of  approximately 0.80 m/s. These conditions result in a chord-based Reynolds number of approximately $3.5\times 10^4$. 

Each rotor has two blades with strut end-plates to minimize parasitic losses~\cite{strom2018jrse}, a height of $H=0.23$ m, diameter of $D=0.172$ m, chord length of $c=0.04$ m, and symmetric NACA0018 blade profile. The blades were pitched radially outward about the quarter chord at a static angle of 6$^\circ$. This turbine geometry has been used in previous studies of single turbines~\cite{polagye2019jrse,hunt2020jrse,scherl2020jrse}. For all cases, the center-to-center separation of the rotors is 1.1D. Consequently, rotor-to-rotor separation distance was $0.1D$ and rotor-to-wall separation distance was approximately $1.2D$.

The turbine rotation rates and relative azimuthal positions 
were regulated by servomotors~\cite{polagye2019jrse} (Yaskawa SGMCS-05B with Yaskawa SGDV-05B3C41 drive). The motors had an internal encoder with over one million counts per revolution and were each fixed to a six-axis load cell (ATI Mini45) that measured all reaction forces and torques and was rigidly mounted to the flume structure. Each turbine driveline consisted of a 12.7 mm diameter stainless steel shaft that terminated, at the lower end, in a bearing that was attached to a second six-axis load cell (ATI Mini45) to measure any parasitic torques present.  This assembly was fixed to the bottom of the flume using a suction plate and scroll vacuum pump (Agilent IDP3) and has been used extensively in previous work~\cite{scherl2020jrse, strom2017natenergy, strom2018jrse}. For performance characterization, the freestream flow velocity was measured 5$D$ upstream at the array mid-plane using an acoustic Doppler velocimeter (Nortek Vectrino profiler).  

\subsection{Flow Field Measurements}

The flow fields acquired using PIV are comprised of two-components of velocity, streamwise and cross-stream in a 2\-D plane downstream of the turbines normal to the blade span, as shown in \Fref{exp1}b. 

The laser used to illuminate the flow was a dual-cavity 30mJ per pulse Nd:YLF laser from Continuum Terra PIV. This laser produced a light sheet with 2 mm thickness. The camera used to acquire the image pairs was a high-speed Vision Research Phantom v641 with a resolution of 2560 x 1600 pixels and calibration of 8.94 pixels/mm resulting in a field-of-view size of 1.66 x 1.04 turbine diameters. The camera was equipped with a 50 mm lens at an F\# of 2. The flow was seeded with 10 $\mu$m hollow-glass beads. DaVis LaVision (version 10.1.1) was used for post-processing the images. 
Background subtraction was done using a Butterworth filter to mitigate residual reflections and background illumination variation. 
Masking was performed by-hand for each case at each azimuthal position where the blade entered the frame. Cross-correlation utilized a multi-grid, multi-pass algorithm with adaptive image deformation, resulting in a final window size of 32 x 32 pixels with a 75\% overlap with 45 vectors per chord length. Spurious vectors were removed with a universal outlier median filter. 

The PIV data field ranged over $X/D \approx 0.4-3$ and $Y/D \approx -1.6 - 1.6$ and was located at the midspan of the turbine blades. Due to the limitations in laser intensity and camera resolution, the data was acquired in a two-by-three grid (\fref{exp1}b). Motorized rails were used to precisely translate cameras in the cross-stream direction. To acquire the downstream fields of view, the turbine assemblies were translated upstream.
The PIV data acquisition was phase-locked with the turbine rotation and triggered when Turbine 1 was pointing directly upstream ($\theta_1 = 0\degree$). Subsequently, 34 image pairs were taken over the course of a full Turbine 1 rotation for 38 full rotations. 
To produce the composite fields, the phase-averaged fields from the six, individual phase-locked acquisitions were stitched together by interpolating to a common grid and performing a weighted average for the overlapped regions. 

Given the data intensive nature of PIV acquisition and processing, the wakes were measured for a subset of phase differences. For both counter-rotating cases, the phases measured were $\phi = 0\degree,\: 45\degree,$ and $90 \degree$. For the co-rotating cases, the phases measured were $\phi = 0\degree,\: 45\degree,\: 90\degree,$ and $\:135 \degree$. Due to the inherent symmetry of the two-bladed turbine array, the counter-rotating $\phi = 135\degree$ cases were not measured because they are identical to the  $\phi = 45\degree$ cases, when mirrored about the $X/D$ axis. 
Vorticity fields were calculated from the streamwise and cross-stream velocity fields using a second-order difference method~\cite{adrian_2011}.

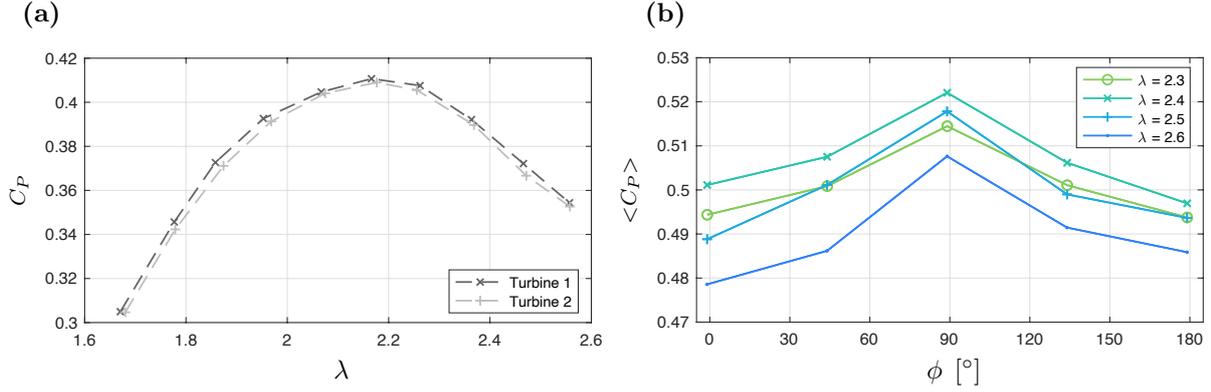
\begin{figure}[t]
\vspace{0.1in}
\centering
\begin{overpic}[width=0.99\textwidth]{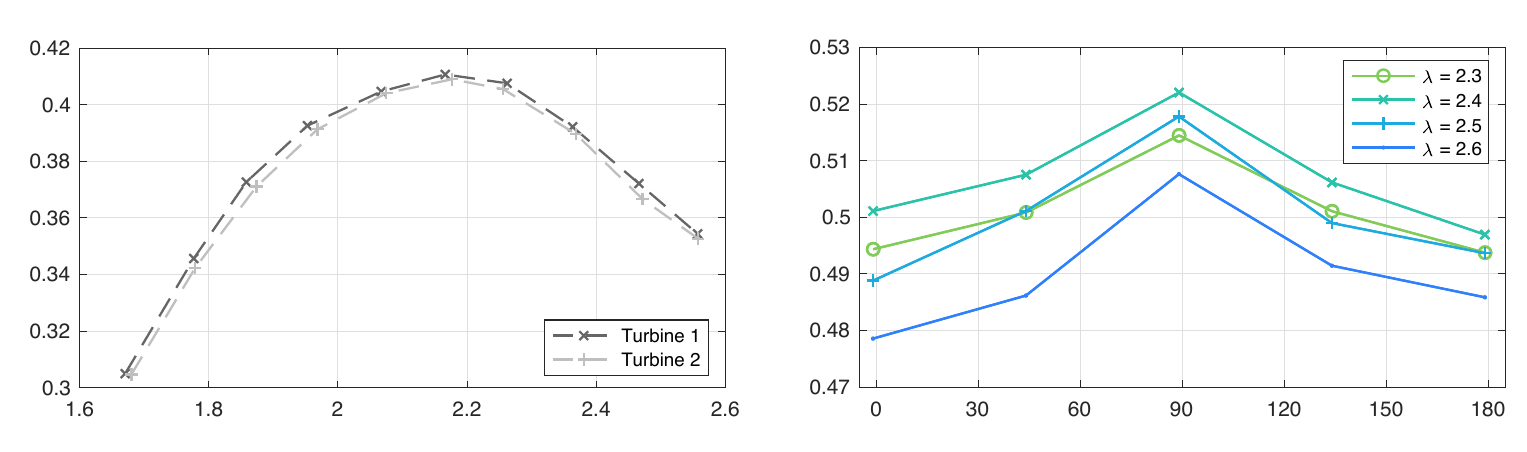}
\put(75.5,-0.5){\small{$\phi \:\:[\degree]$}}
\put(26,-0.5){\small{$\lambda$}}
\put(50,12.5){\begin{sideways}$\scriptstyle{<C_{P}>}$\end{sideways}}
\put(-1,14){\begin{sideways}$\scriptstyle{C_{P}}$\end{sideways}}
\put(0,29.5){\small{\textbf{(a)}}}
\put(52,29.5){\small{\textbf{(b)}}}

\end{overpic}
\vspace{-.07in}
\caption{(a) Characteristic performance ($C_P - \lambda$ curve) for each turbine in isolation (b) Array-level $<C_P> - \phi$ curve for the retreating counter-rotating case at $\lambda = 2.3- 2.6$}\label{fig:cp_phi}
\end{figure}

\subsection{Array Performance}
Individual turbine performance is shown in \Fref{cp_phi}(a) and demonstrates the expected agreement in performance between the two turbine assemblies across tip-speed ratios when they are operated individually. We attribute the slight variations to minor differences in turbine assembly and/or cross-channel inflow variability. 

\begin{figure} [!b]
\centering
\begin{overpic}[width=0.95\textwidth]{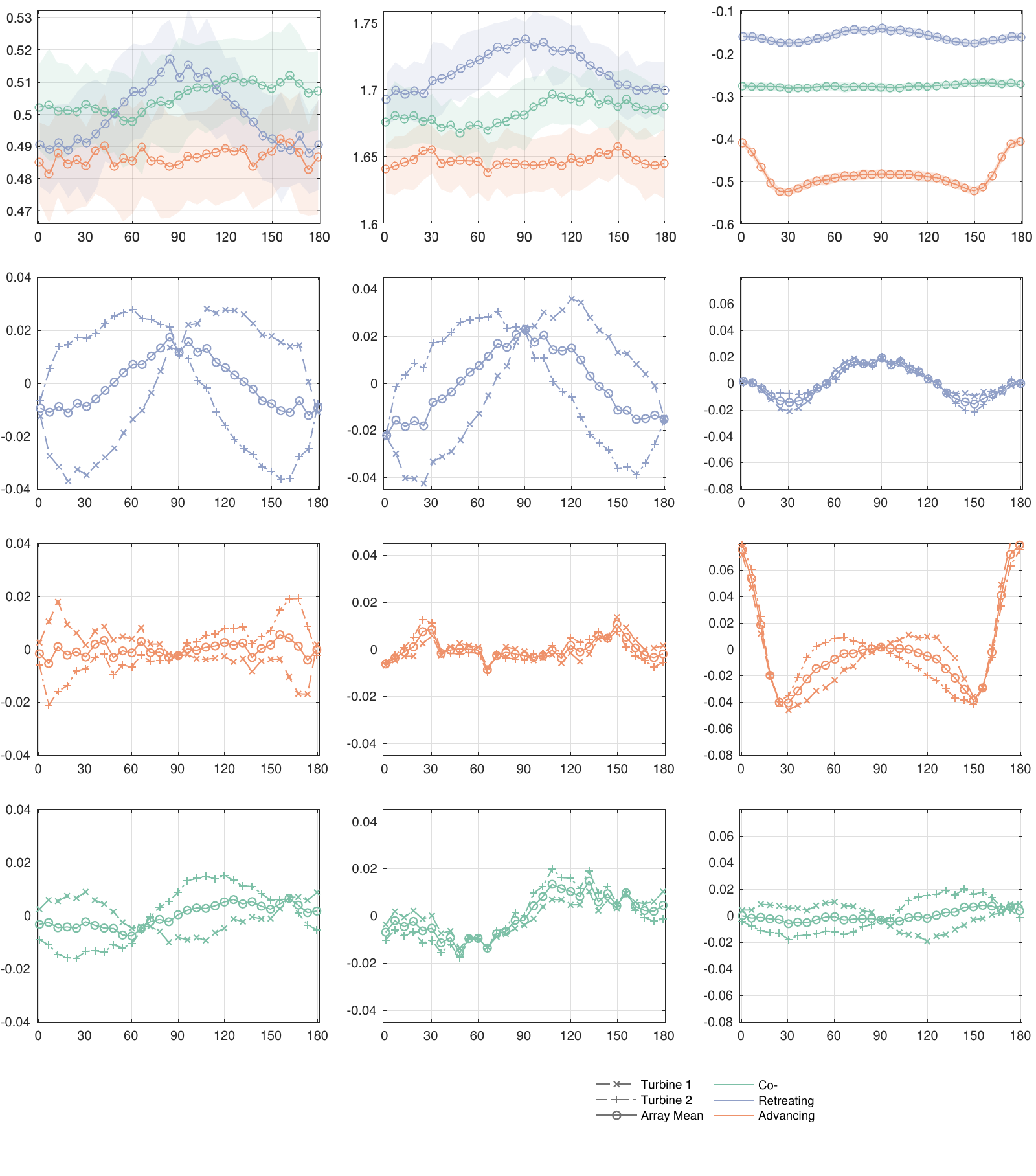}
\put(13,7){\small{$\phi \:\:[\degree]$}}
\put(42.5,7){\small{$\phi \:\:[\degree]$}}
\put(72.75,7){\small{$\phi \:\:[\degree]$}}

\put(-1.5,87.5){\begin{sideways}$\scriptstyle{<C_{P}>}$\end{sideways}}
\put(-1.5,18.5){\begin{sideways}$\scriptstyle{\Delta C_{P}}$\end{sideways}}
\put(-1.5,41.5){\begin{sideways}$\scriptstyle{\Delta C_{P}}$\end{sideways}}
\put(-1.5,65){\begin{sideways}$\scriptstyle{\Delta C_{P}}$\end{sideways}}

\put(57.75,87){\begin{sideways}$\scriptstyle{<C_{F,y}>}$\end{sideways}}
\put(57.75,18.25){\begin{sideways}$\scriptstyle{\Delta C_{F,y}}$\end{sideways}}
\put(57.75,41.25){\begin{sideways}$\scriptstyle{\Delta C_{F,y}}$\end{sideways}}
\put(57.75,64.75){\begin{sideways}$\scriptstyle{\Delta C_{F,y}}$\end{sideways}}

\put(28.25,87){\begin{sideways}$\scriptstyle{<C_{F,x}>}$\end{sideways}}
\put(28.25,18.25){\begin{sideways}$\scriptstyle{\Delta C_{F,x}}$\end{sideways}}
\put(28.25,41.25){\begin{sideways}$\scriptstyle{\Delta C_{F,x}}$\end{sideways}}
\put(28.25,64.75 ){\begin{sideways}$\scriptstyle{\Delta C_{F,x}}$\end{sideways}}

\put(6,100){Performance ($C_P$)}
\put(37.5,100){Thrust ($C_{F,x}$)}
\put(68,100){Lateral ($C_{F,y}$)}

\end{overpic}
\vspace{-.4in}
\caption{Mean array performance, thrust force, and lateral force coefficients for all rotation schemes (top row), mean-subtracted performance, thrust force, and lateral force coefficients for the retreating counter-rotating (second row), advancing counter-rotating (third row), and co-rotating (fourth row) for the individual rotors and array.}

\label{fig:forces_all}
\end{figure}
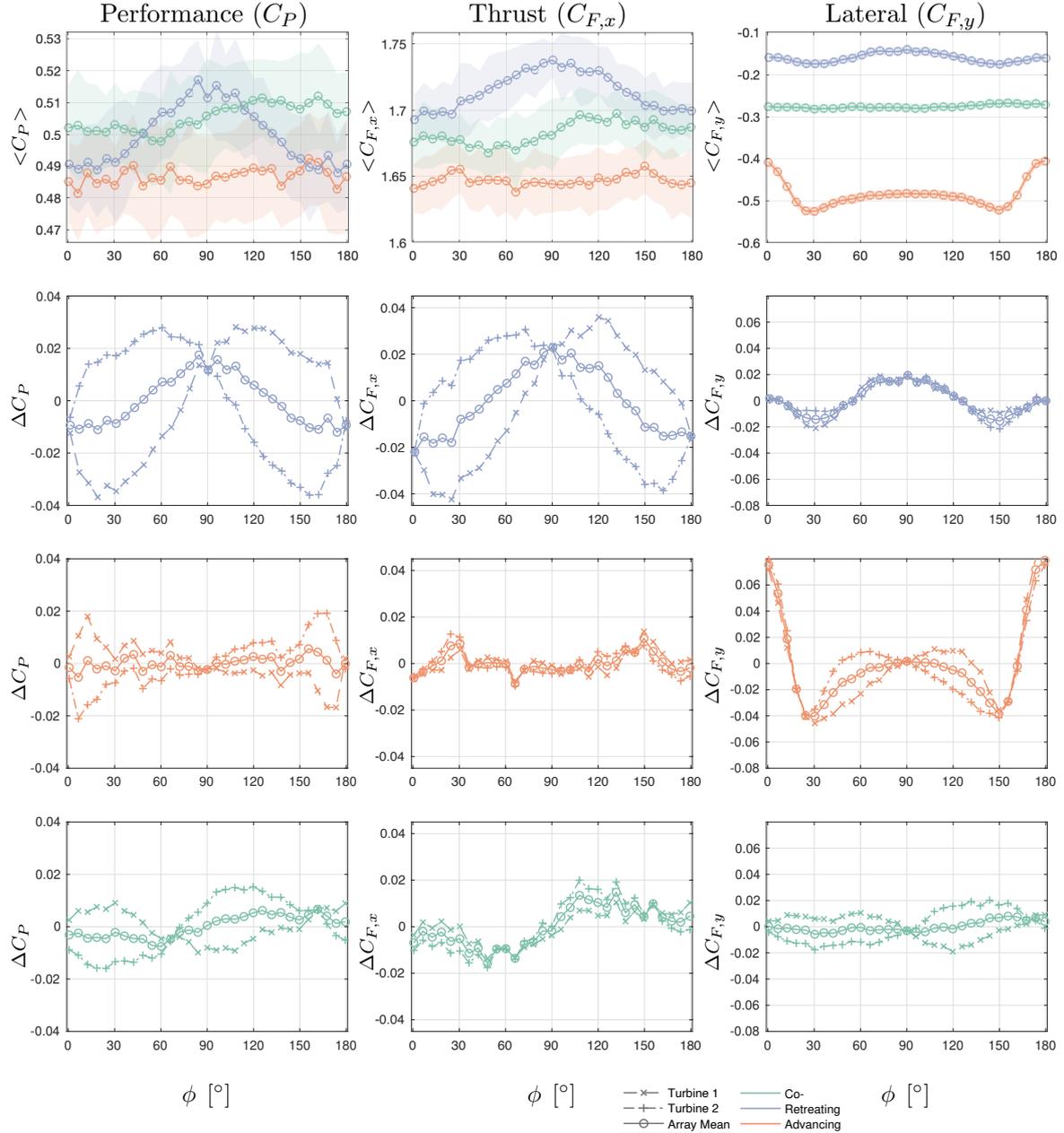

To determine the optimal array tip-speed ratio for the PIV data acquisition, we tested a coarse grid of phase differences at several tip-speed ratios (\Fref{cp_phi}). For these experiments, phase differences were tested in the range of $\phi = 0-180\degree$ in increments of 45$\degree$ for the retreating counter-rotating case. Given the inherent symmetry of a two-bladed cross-flow turbine, the $\phi = 180-360\degree$ range is identical to the tested range. The increase in optimal tip-speed ratio and a portion of the performance between the individual turbines and the array is due to the change in channel blockage between the two cases~\cite{ross2020re}. Blockage, or the ratio of the projected area of the rotor divided by the channel area, tends to increase the optimal tip-speed ratio in addition to increasing the turbine performance~\cite{whelan2009free,ross2020re}. An individual turbine occupies approximately 10\% of the channel cross-section and its peak efficiency is at a tip-speed ratio of $\lambda = 2$, while the array occupies 20\% and the highest average array performance occurs at $\lambda = 2.4$. Consequently, all subsequent array experiments employ $\lambda = 2.4$ for wake measurements and fine-resolution performance measurements across phases.

\section{Results and Discussion}
\subsection{Array Performance}
The array performance and constituent individual turbine performances are shown in the first column of ~\Fref{forces_all}. The top, left panel shows the array-average performance as a function of phase difference for each rotation scheme with the standard deviations of the cycle-averaged performance indicated by the shaded regions. While the standard deviations do overlap, there are still clear variations among the rotation schemes. The subsequent rows of the first column show mean subtracted performance for individual turbines and the array for each of the rotation schemes as a function of phase difference.
The phase differences affect both individual turbine and array-average performance. These differences are meaningful in absolute terms, but relatively subtle. On the whole, we see increased absolute performance for the retreating counter-rotating case over the advancing counter-rotating and co-rotating cases. One possible mechanism for this performance increase is a bulk flow interaction with the blades.

The retreating counter-rotation scheme has the most pronounced phase dependence for both individual turbines and the array average. Since the counter-rotating cases are equal and opposite about $\phi = 90\degree$ within the bounds of experimental uncertainty, we focus on variations in the $\phi = 0-90\degree$ range. Under this scheme, the optimal performance is at a phase difference around $\phi = 90\degree$ and the $\phi = 0-5\degree$ range is the poorest performing. Averaging the two turbines masks larger, offsetting phase dependencies for each turbine which have a performance peak at approximately $\phi = 60\degree$ and trough at approximately $\phi = 20\degree$.

The co-rotating and advancing counter-rotating cases have a less pronounced phase dependence. The array average for the advancing counter-rotating case is nearly independent of phase. However, for the individual turbines we see pronounced performance variability at a phase difference of approximately $\phi = 15\degree$, with a performance increase for Turbine 1 roughly counter-acted by a performance decrease for Turbine 2. The co-rotation case also has fairly constant array-average performance with a somewhat wider range of inter-turbine variability centered around $\phi \,=\, 30\degree$. 
For both rotation schemes, these individual turbine performance variations demonstrate that the phase-difference invariant average performance at an array level is a consequence of the individual turbine performances balancing one another, not an absence of interaction between turbines.

The contribution of each rotor can be further explored through variations in azimuthally position-averaged performance as a function of phase difference and rotation scheme, as shown in \Fref{delta_cp}. It is important to note that small differences in position-averaged performance, when integrated over all positions, result in meaningful cycle-averaged differences (\Fref{forces_all}). In addition, we note that, because performance measurements are taken for the turbine as a whole, the position-averaged performance is shown in the reference frame of one blade, but includes contributions from the second blade. However, the performance from the upstream blade ($\theta = 90\degree - 180\degree$) is dominant~\cite{snortland2023cycle} and allows us to qualitatively relate measurements to its position. 

Turning to \Fref{delta_cp}, if we first examine the position-averaged performance for each rotor (top row), we see that varying the rotation scheme shifts the magnitude and phase at which maximum and minimum performance occur. For the counter-rotating retreating case, we see that maximum performance occurs later in the cycle for both rotors. This points to a possible delay in the phase at which dynamic stall is occurring relative to the other cases. For the counter-rotating advancing case, there is an earlier and slightly lower magnitude maximum phase peak for both rotors, which suggests that dynamic stall is occurring at an earlier azimuthal position. In contrast, for the co-rotating case, Turbine 2 (the retreating rotor) has a later maximum performance than Turbine 1 indicating an appreciable difference between a power stroke that begins with the blade on the bypass flow side of the array (Turbine 2) versus an adjacent turbine (Turbine 1). In aggregate, individual rotors with blades advancing at the array midline (co-rotating Turbine 1, both rotors for the counter-rotating advancing case), we see an earlier performance peak. Conversely, for individual rotors with blades retreating at the array midline (co-rotating Turbine 2, both rotors for the counter-rotating retreating case), we see a delayed performance peak. In summary, the phase of the maximum performance peak depends on the rotation scheme, but is largely independent of the phase offset between rotors.  

These parallels may be due to bulk flow interactions that vary based on the direction the blade is traveling as it approaches the midline of the array, Additionally, fluid-structure interaction between the blades traveling in close proximity to one another likely impacts the boundary layer stability. Finally, the timing of the dynamic stall process likely differs for the two counter-rotating cases. Further research would be required to reach definitive conclusions on these mechanisms.  %

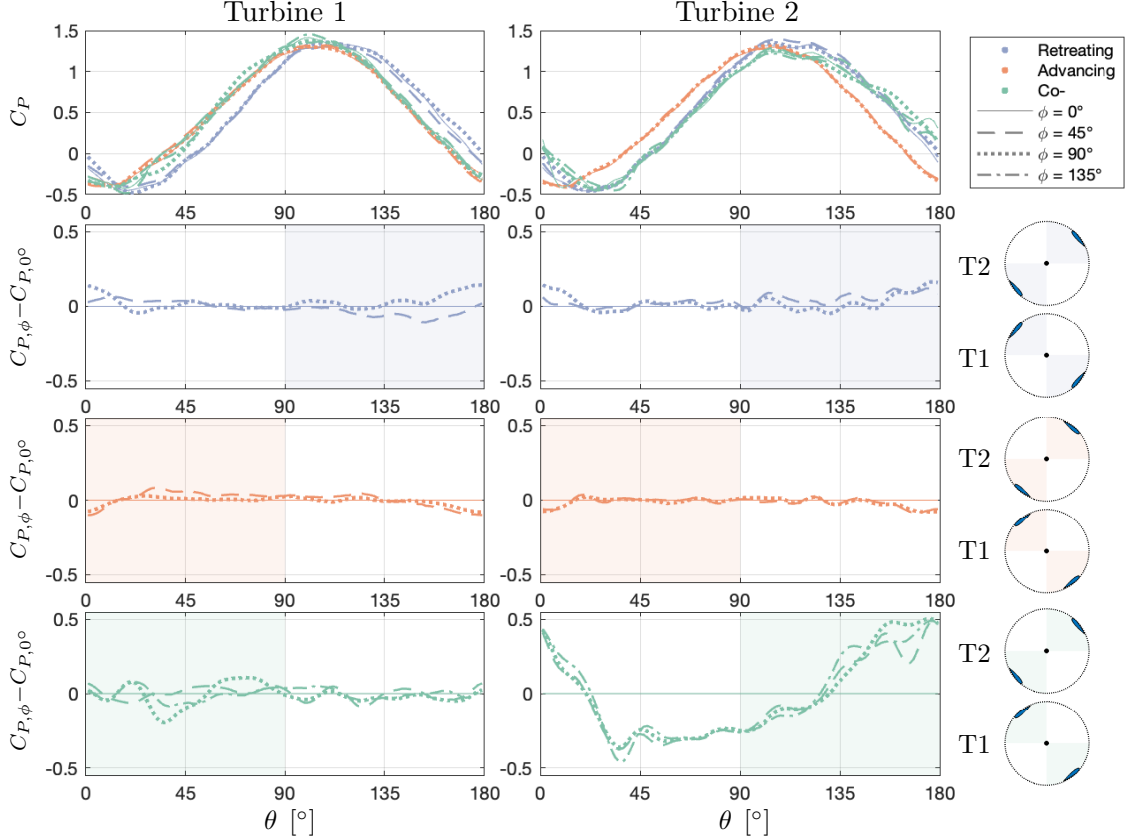
\begin{figure} [t]
\centering
\begin{overpic}[width=0.9\textwidth]{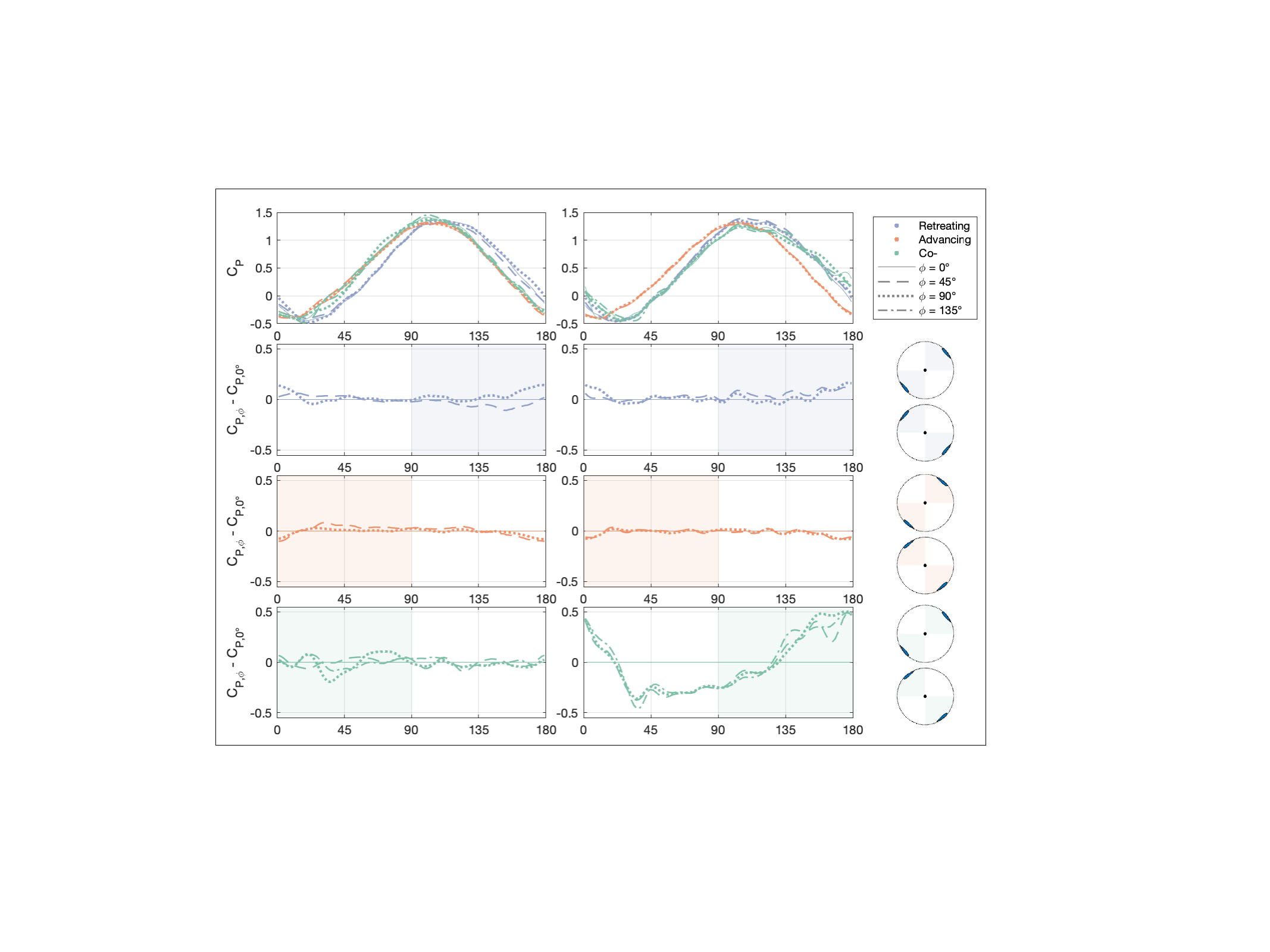}

\put(20.25,-1.75){\small{$\theta \:\:[\degree]$}}
\put(62,-1.75){\small{$\theta \:\:[\degree]$}}
\put(-3,63.25){\begin{sideways}$\scriptstyle{C_P}$\end{sideways}}
\put(-3,41){\begin{sideways}$\scriptstyle{C_{P,\phi}-C_{P,0\degree}}$\end{sideways}}
\put(-3,23.25){\begin{sideways}$\scriptstyle{C_{P,\phi}-C_{P,0\degree}}$\end{sideways}}
\put(-3,5.5){\begin{sideways}$\scriptstyle{C_{P,\phi}-C_{P,0\degree}}$\end{sideways}}
\put(16.5,73){Turbine 1}
\put(58,73){Turbine 2}
\put(84,41){\small{T1}}
\put(84,23.25){\small{T1}}
\put(84,5.25){\small{T1}}

\put(84,49.5){\small{T2}}
\put(84,31.75){\small{T2}}
\put(84,13.75){\small{T2}}
\end{overpic}

\caption{Azimuthal position-averaged performance (top row) for Turbine 1 (left column) and Turbine 2 (right column) for a subset of phase differences relative to the synchronized case ($\phi = 0$) for the counter-rotating retreating case (second row),  counter-rotating advancing case (third row), and co-rotating case (fourth row).}\label{fig:delta_cp}
\end{figure}

Examining the individual rotor performances (second through fourth rows of \Fref{delta_cp}) elucidates some of the variability in performance between rotation schemes in the time averages (\Fref{forces_all}). The performances over azimuthal position for each phase difference are compared by subtracting the synchronized phase difference case ($\phi = 0\degree$). The retreating counter-rotating case (\Fref{delta_cp}, second row) shows that for the $\phi = 45\degree$ and $90\degree$ cases there is a relative increase in performance in the region where one blade is on the inside of the array (between $\theta = 135\degree$ and $25\degree$) for Turbine 2. Whereas at $\phi = 45\degree$, Turbine 1 sees a decrease in performance between $\theta = 90\degree$ and $180\degree$, likely corresponding to the decrease in mean performance (which explains the lobed behavior in~\Fref{forces_all}).
For the advancing counter-rotating rotors (\fref{delta_cp}, third row) there is a slight reduction in performance of the $\phi = 45\degree$ and $90\degree$ at the closest point of approach for both rotors. 
As noted previously, the most variability between rotors is seen in the co-rotating case (\Fref{delta_cp}, third row). For Turbine 1, the performance at each azimuthal position is relatively invariant with phase difference, whereas Turbine 2 has larger performance changes relative to the synchronized case that are somewhat uniform for the $\phi = 45\degree$, = $90\degree$, and $135\degree$ phase differences. 
Here, the rotor that is retreating has increased relative performance in the inter-rotor region attributable to the presence of the second rotor, but then sees a subsequent decrease at intermediate positions.

\subsection{Array Force Variations}
{
Returning to \Fref{forces_all}, we see variations in thrust and lateral forces with control strategy and phase difference. The array-average thrust for each case has a similar profile to performance. The individual rotor thrusts for the retreating counter-rotating case show similar equivalences to performance, but the advancing counter-rotation and co-rotation cases have limited thrust variability. The lateral force variability largely departs from array-average performance and thrust for the counter-rotating schemes, with the advancing counter-rotating case having the strongest phase offset dependence and the retreating case the weakest. For all rotation schemes, individual turbine variability is limited, though appreciable, over a fairly wide range of phase offsets for the advancing counter-rotating case. %
}

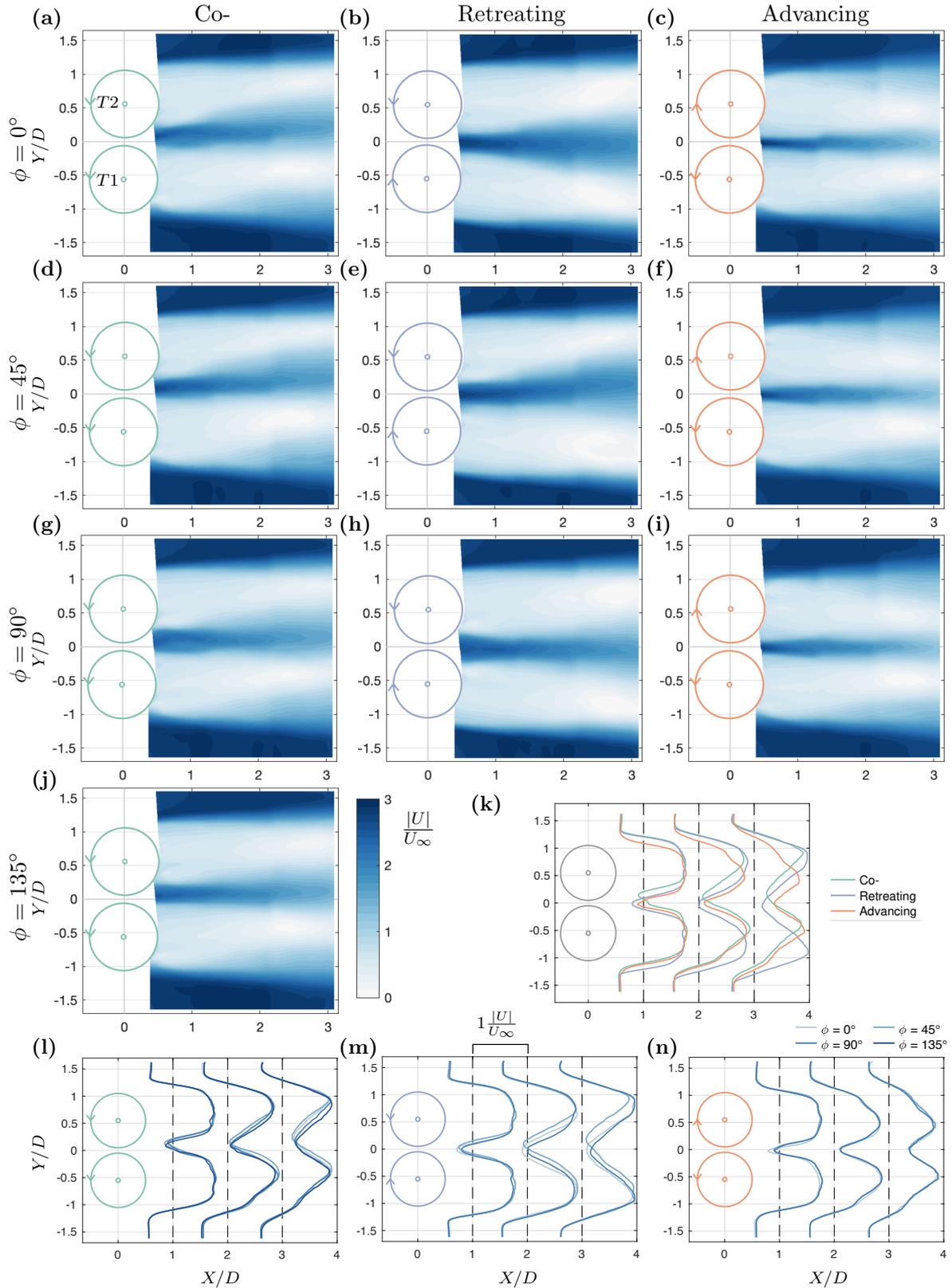
\begin{figure}
\centering
\begin{overpic}[width=0.99\textwidth]{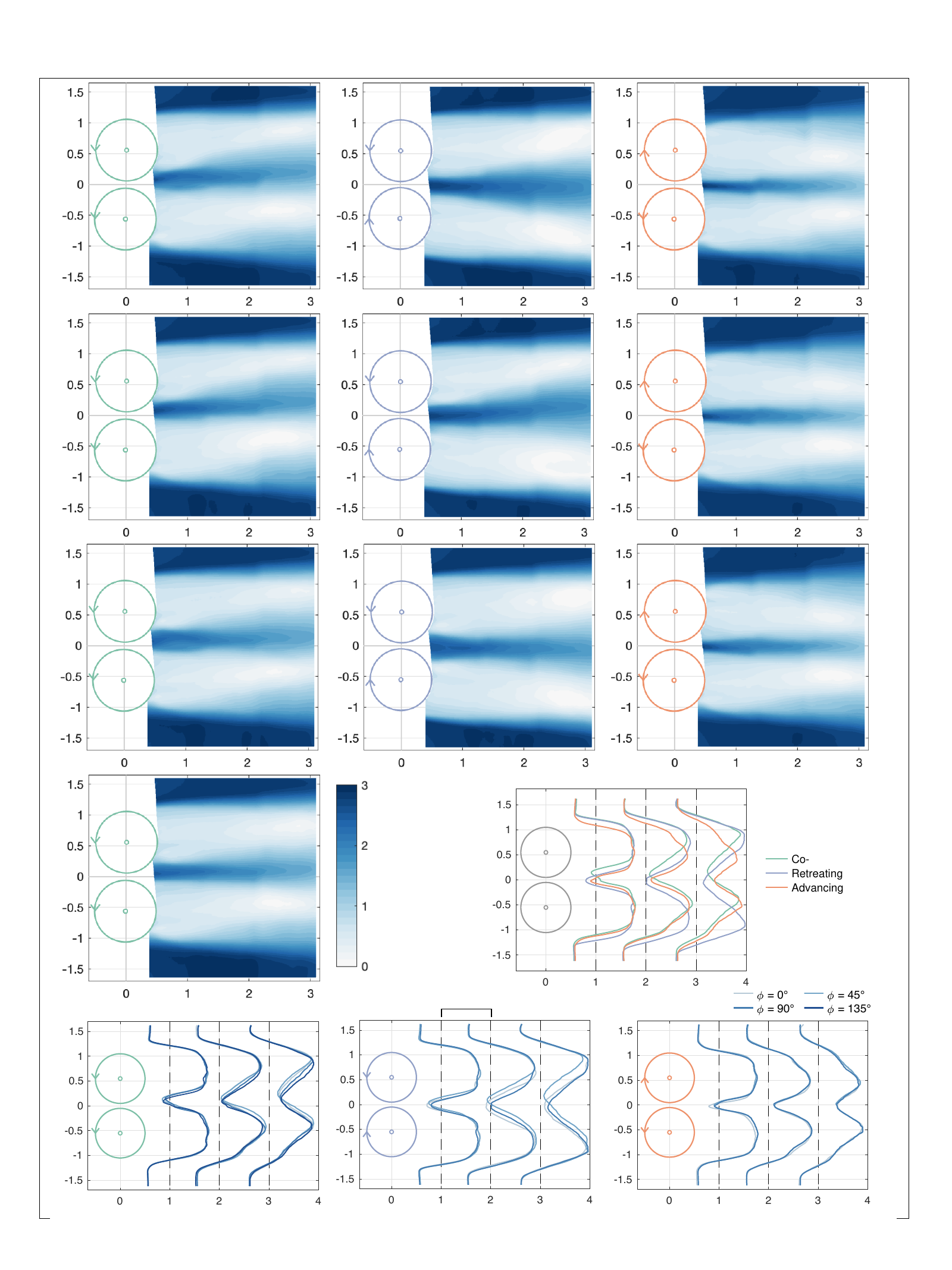}
\put(13.5, -1){$\scriptstyle{X/D}$}
\put(37.25, -1){$\scriptstyle{X/D}$}
\put(62, -1){$\scriptstyle{X/D}$}

\put(0,8){\begin{sideways}$\scriptstyle{Y/D}$\end{sideways}}
\put(0,28.5){\begin{sideways}$\scriptstyle{Y/D}$\end{sideways}}
\put(0,48.5){\begin{sideways}$\scriptstyle{Y/D}$\end{sideways}}
\put(0,69){\begin{sideways}$\scriptstyle{Y/D}$\end{sideways}}
\put(0,89){\begin{sideways}$\scriptstyle{Y/D}$\end{sideways}}

\put(-1.75,26.5){\begin{sideways}${\phi = 135\degree}$\end{sideways}}
\put(-1.75,47.25){\begin{sideways}${\phi = 90\degree}$\end{sideways}}
\put(-1.75,67.5){\begin{sideways}${\phi = 45\degree}$\end{sideways}}
\put(-1.75,87.5){\begin{sideways}${\phi = 0\degree}$\end{sideways}}

\put(13,100.5){Co-}
\put(34,100.5){Retreating}
\put(58.25,100.5){Advancing}

\put(29.5,35){$\frac{|U|}{U_\infty}$}

\put(0,100.25){\small{\textbf{(a)}}}
\put(24.5,100.25){\small{\textbf{(b)}}}
\put(49,100.25){\small{\textbf{(c)}}}

\put(0,80){\small{\textbf{(d)}}}
\put(24.5,80){\small{\textbf{(e)}}}
\put(49,80){\small{\textbf{(f)}}}

\put(0,59.5){\small{\textbf{(g)}}}
\put(24.5,59.5){\small{\textbf{(h)}}}
\put(49, 59.5){\small{\textbf{(i)}}}

\put(0,39){\small{\textbf{(j)}}}
\put(35 ,37){\small{\textbf{(k)}}}

\put(0,17.5){\small{\textbf{(l)}}}
\put(24.5,17.5){\small{\textbf{(m)}}}
\put(49,17.5){\small{\textbf{(n)}}}

\put(35.5,19.5){$\scriptstyle{1\frac{|U|}{U_\infty}}$}

\put(5.2, 93.5){$\scriptstyle{T2}$}
\put(5.2, 87.25){$\scriptstyle{T1}$}

\end{overpic}
\vspace{-.05in}
\caption{(a-j) Mean velocity magnitude normalized by free-stream flow velocity for all combinations of rotation scheme and phase difference. (k) Wake profiles for each rotation scheme at a $\phi = 90\degree$ phase difference. (l-n) Wake profiles for each rotation scheme and phase difference. For each of the profiles, the contour represents the velocity magnitude at $X/D = 1,\; 2, \; or \; 3$. The spacing between the dotted lines (1 $X/D$) represents $1\frac{|U|}{U_\infty}$ and at the dotted lines, the magnitude of the flow is equal to the free-stream($\frac{|U|}{U_\infty}$ = 1).}\label{fig:magdef}
\end{figure}

\subsection{Flow Field Evolution}

The time-average wake velocity magnitude for each of the rotation schemes and the subset of phase differences are shown in \Fref{magdef} a-j and wake profiles at specific cross-sections ($X/D$ = 1, 2, and 3) are shown in \Fref{magdef}k-f. Broadly, each of the wakes has similar structure. As demonstrated by the profiles, they are comprised of (1) a wake deficits that skew moderately towards the advancing side of the respective rotor, (2) moderately accelerated flow at the array mid-line ($Y/D = 0$), and (3) strongly accelerated flow in the bypass regions.  The flow acceleration is due to the bluff-body effects where the flow is diverted around the turbine rather than passing through. This acceleration is augmented by the channel blockage of the array.  In all cases, the largest velocity deficit is at $x/D \approx 2.5$, which is consistent with prior investigations~\cite{araya2017jfm}.

\Fref{magdef}k shows wake profiles for each of the rotation at a phase difference of $\phi = 90$. There are differences in magnitude of the ``impingement region'' (i.e., the accelerated flow between the turbines) and its evolution as the wake mixes and skews. The relative locations of the peaks and troughs of the contours are the clearest indicator of the skew direction and are most evident at $X/D = 3$. The commonalities in general structure across cases can be contrasted with differences between rotation schemes and phase differences. For example, the width, magnitude, and diffusion of the impingement jet, which is nearly independent of rotation direction and phase difference at $X/D$ = 1, notably varies further downstream.

\begin{figure} [b!]
\vspace{0.1in}
\centering
\begin{overpic}[width=0.9\textwidth]{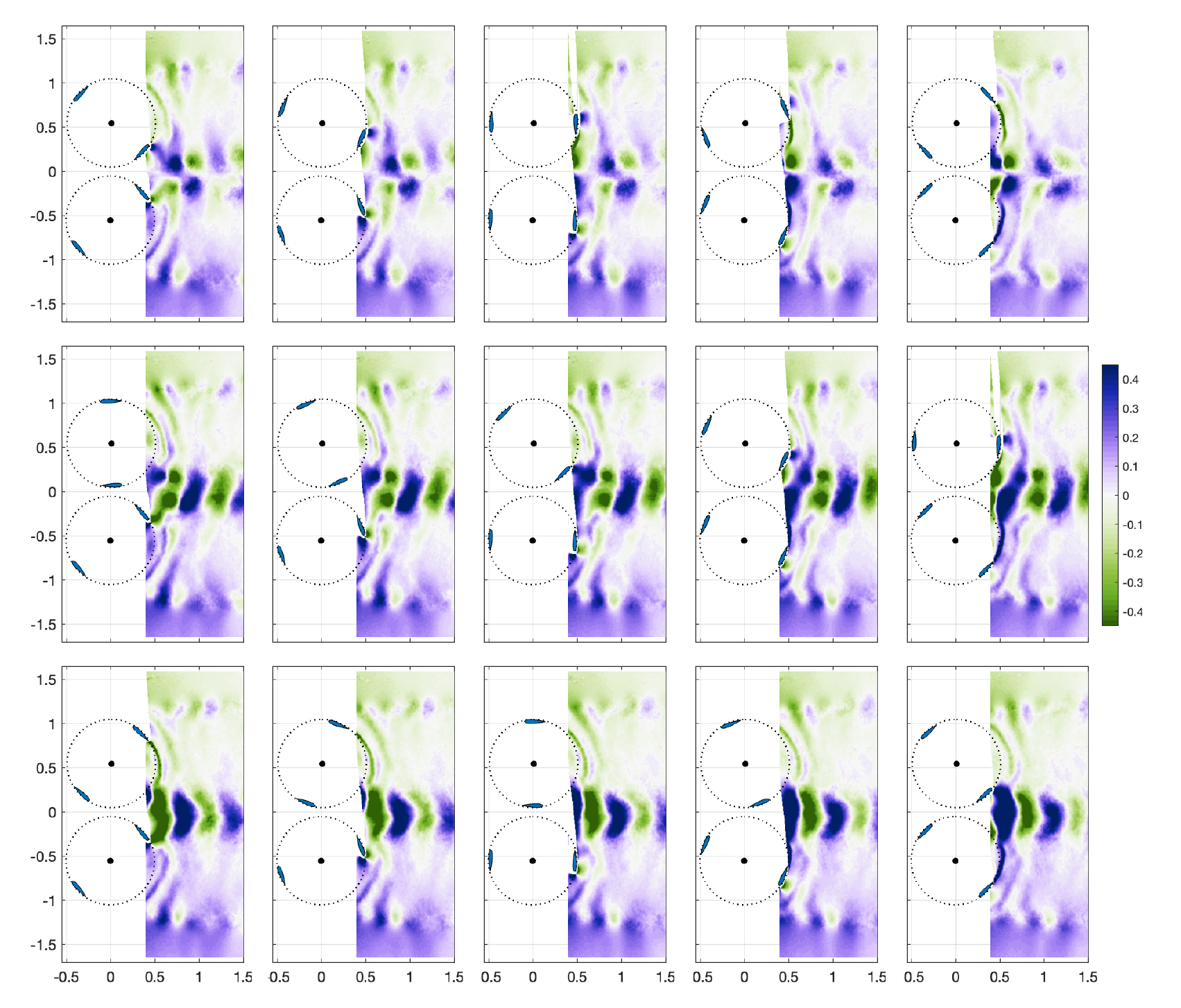}

\put(10.75, 0){$\scriptstyle{X/D}$}
\put(28.5, 0){$\scriptstyle{X/D}$}
\put(46.5, 0){$\scriptstyle{X/D}$}
\put(64.5, 0){$\scriptstyle{X/D}$}
\put(82.25, 0){$\scriptstyle{X/D}$}

\put(-0.5,41.5){\begin{sideways}$\scriptstyle{Y/D}$\end{sideways}}
\put(-4,40){\begin{sideways}$\phi = 45\degree$\end{sideways}}

\put(-0.5,14.5){\begin{sideways}$\scriptstyle{Y/D}$\end{sideways}}
\put(-4,12){\begin{sideways}$\phi = 90\degree$\end{sideways}}

\put(-0.5,69){\begin{sideways}$\scriptstyle{Y/D}$\end{sideways}}
\put(-4,67.5){\begin{sideways}$\phi = 0\degree$\end{sideways}}

\put(9,84){$\theta_1 \approx 43\degree$}
\put(25.5,84){$\theta_1 \approx 64\degree$}
\put(44,84){$\theta_1 \approx 85\degree$}
\put(61.5,84){$\theta_1 \approx 107\degree$}
\put(79.5,84){$\theta_1 \approx 128\degree$}

\put(6.25, 74.5){$\scriptstyle{T2}$}
\put(6.25, 66){$\scriptstyle{T1}$}

\put(98, 42.25){${v}$}
\end{overpic}

\caption{Truncated, position-average cross-stream velocity fields ($v$) for the retreating counter-rotation scheme at $\phi = 0\degree$, $45\degree$, and $90\degree$ at select rotor positions. }\label{fig:vvel}
\end{figure}

The advancing counter-rotating case (\Fref{magdef}c,f,i,n), produces the narrowest jet and also the one that diffuses most rapidly. The wake profile contours (\Fref{magdef}n), show that the wake of each rotor for this case is steered towards $Y/D=0$. In other words, the maximum wake deficit approaches $Y/D=0$ with increasing $X/D$. For this case, the mean wake evolution is independent of phase difference. This uniformity is likely a consequence of relatively limited interaction between the rotors and the absence of direct interaction between the energetic wake structures on the retreating sides of the individual turbines (see further discussion of position-average flow fields). 

For the retreating counter-rotating case, the impingement jet (\Fref{magdef}b,e,h,m) is  widest and diffuses more slowly than the advancing counter-rotating case. The difference in the jet width and magnitude are likely related to the flow entrainment in the direction of blade rotation. In other words, for the retreating counter-rotating case, blades are moving in the same direction as the bulk flow leading to a higher magnitude jet between the rotors. 
This can be contrasted with the lower magnitude jet in the advancing counter-rotating case where the flow entrainment acts opposite the direction of freestream flow. These differences may contribute to the performance differences between the two rotation schemes. 
However, it remains difficult to establish causality between the performance and bulk flow structures in the inter rotor region in part due to the presence of the second blade and in part due to the the close proximity the blades are traveling, likely leading to pressure field alterations and associated boundary layer instability that cannot be observed with these experimental techniques. 
For the retreating counter-rotating profiles (\Fref{magdef}k,m), both turbine wakes skew away from the array midline. Unlike the advancing counter-rotating case, the mean wake evolution depends on the phase difference. At the $\phi = 0\degree$ and $90\degree$ phase differences, the wake evolution is symmetric about the midplane, which is consistent with the inherent cross-stream symmetry of the array. Conversely, in the $\phi = 45\degree$ case, the array wake skews towards the positive $Y/D$ direction. %

The wake evolution in the co-rotating case is an amalgam of the characteristics of the advancing and retreating counter-rotating cases. The magnitude of the impingement jet and its diffusion is similar to the retreating counter-rotating case, but narrower in width. The profiles for the co-rotation case (\Fref{magdef}l) shows both of the wakes skewed in parallel towards the positive $Y/D$ direction. Variation in phase difference produces minor changes in the wake evolution at $X/D > 2$ that are less pronounced than for the retreating counter-rotating case.

\begin{figure}
\vspace{0.1in}
\centering
\begin{overpic}[width=0.9\textwidth]{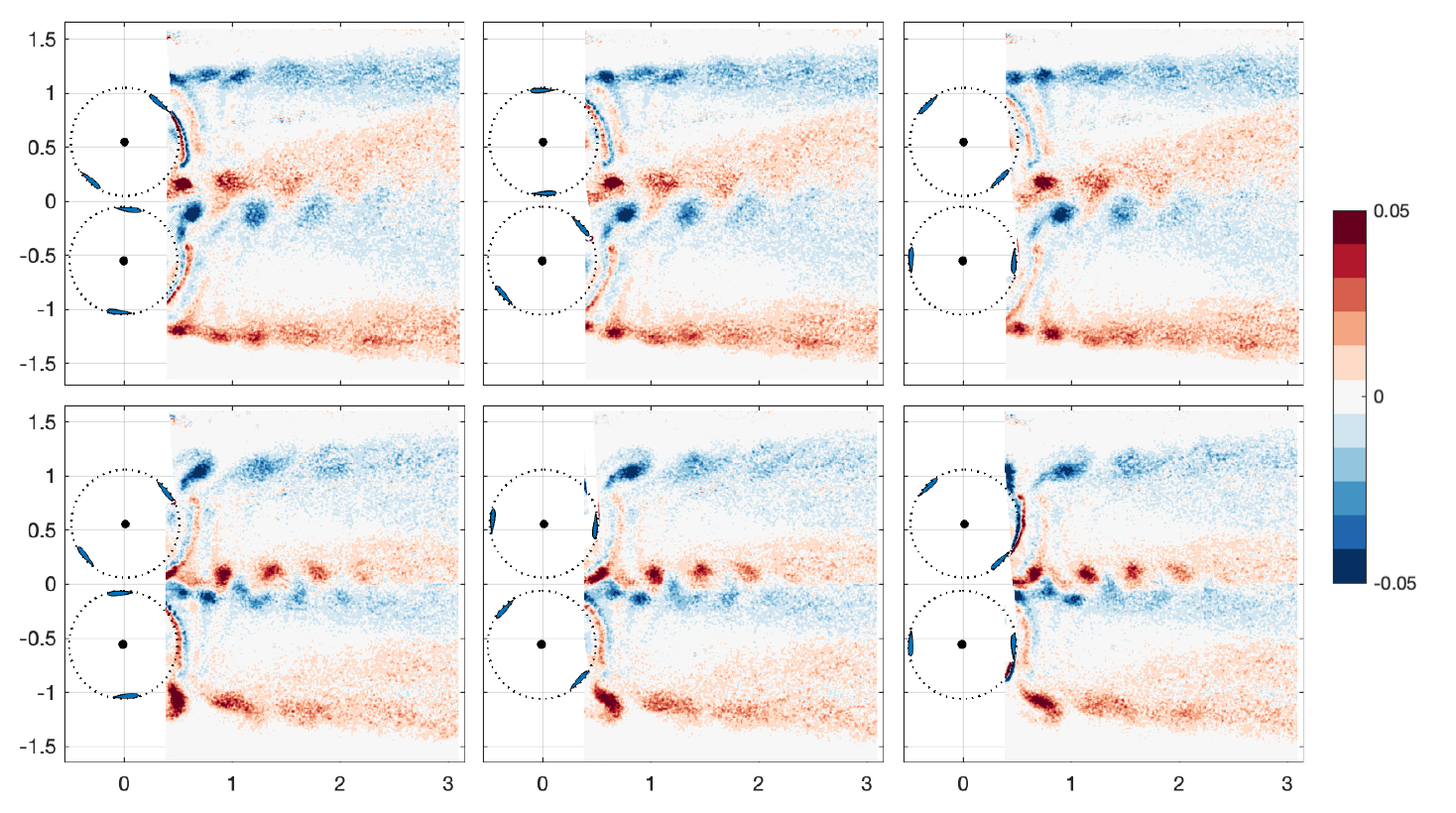}

\put(17.5, -1){$\scriptstyle{X/D}$}
\put(47, -1){$\scriptstyle{X/D}$}
\put(75, -1){$\scriptstyle{X/D}$}
\put(-0.5,39.5){\begin{sideways}$\scriptstyle{Y/D}$\end{sideways}}
\put(-0.5,12.5){\begin{sideways}$\scriptstyle{Y/D}$\end{sideways}}

\put(-4,33.5){\begin{sideways}Retreating\end{sideways}}
\put(-4,8.5){\begin{sideways}Advancing\end{sideways}}

\put(100, 30){$\scriptstyle{\omega}$}

\put(14.5,55.5){$\theta_1 \approx 0\degree$}
\put(44,55.5){$\theta_1 \approx 42\degree$}
\put(72.5,55.5){$\theta_1 \approx 84\degree$}

\put(5.2, 5){$\scriptstyle{T2}$}
\put(5.2, 5){$\scriptstyle{T1}$}

\end{overpic}

\caption{Advancing and retreating counter-rotating vorticity fields for the $\phi = 45\degree$ phase difference case.}\label{fig:frames}
\end{figure}
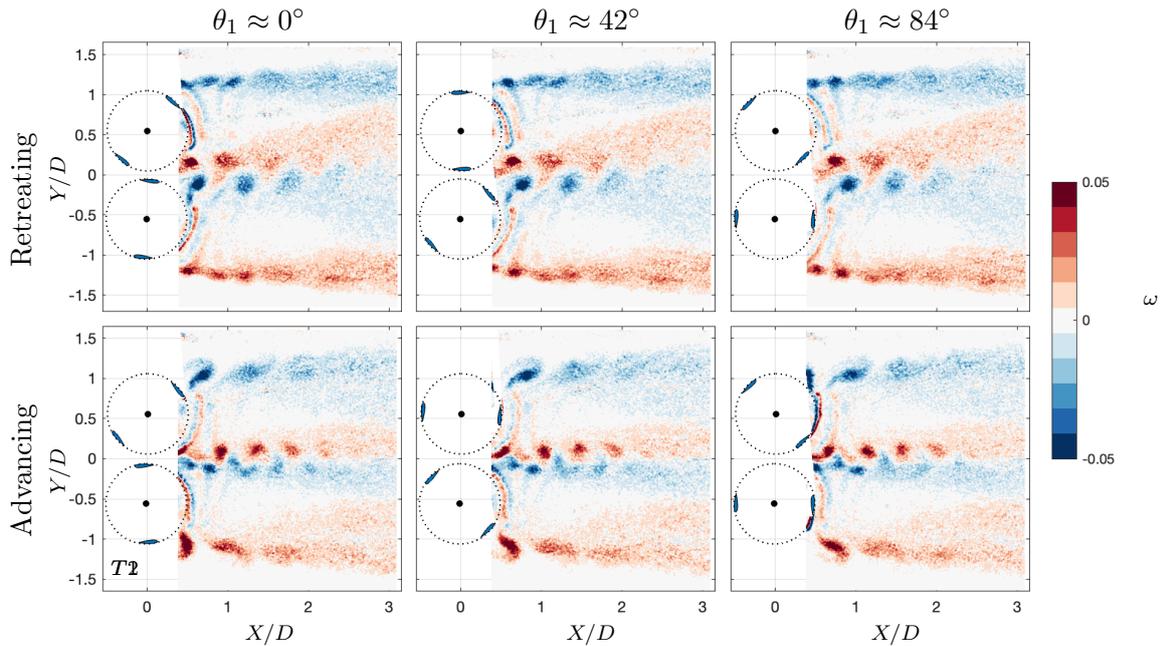

\Fref{frames} shows vorticity at three blade positions for the two counter-rotating cases at a phase difference of $\phi = 45\degree$. This highlights alterations in the vortex interactions between rotation schemes. For both cases, vortices persist further downstream in the inside of the array, even though the dynamic stall vortex shed at the conclusion of the upstream sweep alternates between the array midline and bypass flow. This persistence is likely due to a stronger shear layer between the rotor wake and bypass flow, inducing stronger mixing. In other words, as for performance, there is a clear difference between the rotor-wall and rotor-rotor wake interactions. %
Additionally, for the retreating counter-rotating case, the larger structures at the array midline could be entraining additional flow causing the wider jet evident in the flow magnitude plots (~\Fref{magdef}). %

The skew towards $Y/D > 0$ in the time-average wake for the retreating counter-rotating case is apparent  in \Fref{frames}. This is explored in more detail in \Fref{vvel}, which shows the cross-stream velocity ($v$) for the retreating counter-rotation cases at discrete rotor positions at all three phase differences ($\phi = 0\degree,\,45\degree,\,\& \,90\degree$). When there is no phase difference between the rotors (top row, $\phi = 0\degree$) at the midline of the array the coherent structures from each rotor are shed simultaneously and interact as they advect in the impingement jet. For the $\phi = 45\degree$ phase difference (middle row), we see that the positive cross-stream velocity region intercepts the blade that follows from Turbine 2 at $\theta_1 = 85\degree$, which is traveling in the $+y$-direction. This interaction appears to augment this positive cross-stream velocity region and only occurs because the Turbine 2 blade is closely following the Turbine 1 blade. This interaction is unique to the counter-rotation retreating-$\phi = 45\degree$ difference case and shows a possible mechanism for the $+y$-direction wake skew. 
At a $\phi = 90\degree$ phase offset, we see the positive and negative velocity regions alternate as they are each shed from their corresponding rotors. For both the $90\degree$ and $0\degree$ cases all flow structures are symmetric and no skew is present.

\section{Conclusions}

In this work, we present comprehensive performance, force, and wake measurements for three configurations of cross-flow turbine pairs. For each rotation scheme, the turbines are tested across phase differences. We find that the main driver of performance, forces, and wake is the rotation scheme, with secondary dependencies on phase differences. We find that rotor-rotor interactions have a significant effect on performance, particularly in the inter-rotor region. For each rotation scheme, the dominant dynamics persist across phase differences. The scheme in which counter-rotating blades are retreating or traveling downstream at their closest point of approach has the largest phase-difference dependent variations in performance and wake dynamics. 
For the cases of counter-rotating blades advancing at their closest point of approach or co-rotating blades, there are only small phase perturbations in time-average array performance and wake evolution with phase difference. However, phase variations do contribute to varied power outputs and forces from individual turbines that are balanced out in aggregate at the array level. For the mean wakes, consistent with previous work, we find that the individual rotor wakes skew towards the advancing side of the rotor which causes a divergent, convergent, and parallel wake for the retreating, advancing, and co-rotation cases, respectively.

These results could be used to design rotor arrays through wake steering. To design arrays with multiple rows it would be important to consider the array wake skew. The convergent wake dynamics in the advancing counter-rotating case could be worthwhile despite the decreased array performance because of the narrower wake. If there is only one row in an array, from a performance standpoint it would likely be advantageous to have pairs of rotors operating under the counter-rotating retreating scheme. Additionally, regardless of whether or not phase offset is controlled for, a benefit may be seen for side-by-side rotors when blades are retreating. 

Drawing connections between wake evolution and performance variations remains speculative. Future work utilizing measurements within the rotors and resolving forces on individual blades could illuminate some of these dependencies. Additionally, wake dynamics could be further explored to understand more of the time- and position-dependent fluid mechanisms present and how they change with rotation scheme and phase difference. Three-dimensional effects could also be examined by measuring spanwise velocity, as well as collecting tomographic measurements to understand edge effects in the wake and their influence on mid-plane wake evolution.

\bibliography{PIV.bib}

\begin{thebibliography}{10}

\bibitem{adrian_2011}
Ronald~J Adrian and Jerry Westerweel.
\newblock {\em Particle image velocimetry}.
\newblock Number~30. Cambridge University Press, 2011.

\bibitem{ahmadi_2016}
Mojtaba Ahmadi-Baloutaki, Rupp Carriveau, and David~SK Ting.
\newblock A wind tunnel study on the aerodynamic interaction of vertical axis
  wind turbines in array configurations.
\newblock {\em Renewable Energy}, 96:904--913, 2016.

\bibitem{araya2017jfm}
Daniel~B Araya, Tim Colonius, and John~O Dabiri.
\newblock Transition to bluff-body dynamics in the wake of vertical-axis wind
  turbines.
\newblock {\em Journal of Fluid Mechanics}, 813:346--381, 2017.

\bibitem{athair2023intracycle}
Ari Athair, Abigale Snortland, Isabel Scherl, Brian Polagye, and Owen Williams.
\newblock Intracycle control sensitivity of cross-flow turbines.
\newblock In {\em Proceedings of the European Wave and Tidal Energy
  Conference}, volume~15, 2023.

\bibitem{bachant2015jot}
Peter Bachant and Martin Wosnik.
\newblock Characterising the near-wake of a cross-flow turbine.
\newblock {\em Journal of Turbulence}, 16(4):392--410, 2015.

\bibitem{bachant2016energies}
Peter Bachant and Martin Wosnik.
\newblock Effects of reynolds number on the energy conversion and near-wake
  dynamics of a high solidity vertical-axis cross-flow turbine.
\newblock {\em Energies}, 9(2):73, 2016.

\bibitem{bachant2016plos}
Peter Bachant, Martin Wosnik, Budi Gunawan, and Vincent~S Neary.
\newblock Experimental study of a reference model vertical-axis cross-flow
  turbine.
\newblock {\em PloS one}, 11(9):e0163799, 2016.

\bibitem{bremseth_2016}
J~Bremseth and K~Duraisamy.
\newblock Computational analysis of vertical axis wind turbine arrays.
\newblock {\em Theoretical and Computational Fluid Dynamics}, 30(5):387--401,
  2016.

\bibitem{brownstein2016jrse}
Ian~D Brownstein, Matthias Kinzel, and John~O Dabiri.
\newblock Performance enhancement of downstream vertical-axis wind turbines.
\newblock {\em Journal of Renewable and Sustainable Energy}, 8(5):053306, 2016.

\bibitem{brownstein2019energies}
Ian~D Brownstein, Nathaniel~J Wei, and John~O Dabiri.
\newblock Aerodynamically interacting vertical-axis wind turbines: Performance
  enhancement and three-dimensional flow.
\newblock {\em Energies}, 12(14):2724, 2019.

\bibitem{carrigan_2012}
Travis~J Carrigan, Brian~H Dennis, Zhen~X Han, and Bo~P Wang.
\newblock Aerodynamic shape optimization of a vertical-axis wind turbine using
  differential evolution.
\newblock {\em ISRN Renewable Energy}, 2012, 2012.

\bibitem{castelli_2012}
Marco~Raciti Castelli and Ernesto Benini.
\newblock Effect of blade inclination angle on a darrieus wind turbine.
\newblock {\em Journal of turbomachinery}, 134(3):031016, 2012.

\bibitem{chen_2013}
Chein-Chang Chen and Cheng-Hsiung Kuo.
\newblock Effects of pitch angle and blade camber on flow characteristics and
  performance of small-size darrieus vawt.
\newblock {\em Journal of Visualization}, 16(1):65--74, 2013.

\bibitem{chen_2017}
Wei-Hsin Chen, Ching-Ying Chen, Chun-Yen Huang, and Chii-Jong Hwang.
\newblock Power output analysis and optimization of two straight-bladed
  vertical-axis wind turbines.
\newblock {\em Applied energy}, 185:223--232, 2017.

\bibitem{corke2015ar}
Thomas~C Corke and Flint~O Thomas.
\newblock Dynamic stall in pitching airfoils: aerodynamic damping and
  compressibility effects.
\newblock {\em Annual Review of Fluid Mechanics}, 47:479--505, 2015.

\bibitem{dabiri2011jrse}
John~O Dabiri.
\newblock Potential order-of-magnitude enhancement of wind farm power density
  via counter-rotating vertical-axis wind turbine arrays.
\newblock {\em Journal of renewable and sustainable energy}, 3(4):043104, 2011.

\bibitem{dave2021jrse}
Mukul Dave and Jennifer~A Franck.
\newblock Comparison of rans and les for a cross-flow turbine in confined and
  unconfined flow.
\newblock {\em Journal of Renewable and Sustainable Energy}, 13(6):064503,
  2021.

\bibitem{dave2023analysis}
Mukul Dave and Jennifer~A Franck.
\newblock Analysis of dynamic stall development on a cross-flow turbine blade.
\newblock {\em Physical Review Fluids}, 8(7):074702, 2023.

\bibitem{dave2021aiaa}
Mukul Dave, Benjamin Strom, Abigale Snortland, Owen Williams, Brian Polagye,
  and Jennifer~A Franck.
\newblock Simulations of intracycle angular velocity control for a crossflow
  turbine.
\newblock {\em AIAA Journal}, 59(3):812--824, 2021.

\bibitem{de2018towards}
D~De~Tavernier, Carlos Ferreira, Ang Li, US~Paulsen, and HA~Madsen.
\newblock Towards the understanding of vertical-axis wind turbines in
  double-rotor configuration.
\newblock In {\em Journal of Physics: Conference Series}, volume 1037, page
  022015. IOP Publishing, 2018.

\bibitem{doan2020jmse}
Minh~N Doan, Yuriko Kai, and Shinnosuke Obi.
\newblock Twin marine hydrokinetic cross-flow turbines in counter rotating
  configurations: A laboratory-scaled apparatus for power measurement.
\newblock {\em Journal of Marine Science and Engineering}, 8(11):918, 2020.

\bibitem{dunne2015eif}
Reeve Dunne and Beverley~J McKeon.
\newblock Dynamic stall on a pitching and surging airfoil.
\newblock {\em Experiments in Fluids}, 56(8):157, 2015.

\bibitem{duraisamy_2014}
Karthikeyan Duraisamy and Vinod Lakshminarayan.
\newblock Flow physics and performance of vertical axis wind turbine arrays.
\newblock In {\em 32nd AIAA Applied Aerodynamics Conference}, page 3139, 2014.

\bibitem{durrani_2011}
Naveed Durrani, Ning Qin, Harris Hameed, and Shahab Khushnood.
\newblock 2-d numerical analysis of a vawt wind farm for different
  configurations.
\newblock In {\em 49th AIAA Aerospace Sciences Meeting including the New
  Horizons Forum and Aerospace Exposition}, page 461, 2011.

\bibitem{ellington1984aerodynamics}
Charles~Porter Ellington.
\newblock The aerodynamics of hovering insect flight. i. the quasi-steady
  analysis.
\newblock {\em Philosophical Transactions of the Royal Society of London. B,
  Biological Sciences}, 305(1122):1--15, 1984.

\bibitem{eriksson2008jrse}
Sandra Eriksson, Hans Bernhoff, and Mats Leijon.
\newblock Evaluation of different turbine concepts for wind power.
\newblock {\em renewable and sustainable energy reviews}, 12(5):1419--1434,
  2008.

\bibitem{fish2006ar}
FE~Fish and GV~Lauder.
\newblock Passive and active flow control by swimming fishes and mammals.
\newblock {\em Annu. Rev. Fluid Mech.}, 38:193--224, 2006.

\bibitem{gauvin2022}
Olivier Gauvin-Tremblay and Guy Dumas.
\newblock Hydrokinetic turbine array analysis and optimization integrating
  blockage effects and turbine-wake interactions.
\newblock {\em Renewable Energy}, 181:851--869, 2022.

\bibitem{gauvin2022hydrokinetic}
Olivier Gauvin-Tremblay and Guy Dumas.
\newblock Hydrokinetic turbine array analysis and optimization integrating
  blockage effects and turbine-wake interactions.
\newblock {\em Renewable Energy}, 181:851--869, 2022.

\bibitem{huang2023experimental}
Ming Huang, Yugandhar Vijaykumar~Patil, Andrea Sciacchitano, and Carlos
  Ferreira.
\newblock Experimental study of the wake interaction between two vertical axis
  wind turbines.
\newblock {\em Wind Energy}, 2023.

\bibitem{hunt2020jrse}
Aidan Hunt, Carl Stringer, and Brian Polagye.
\newblock Effect of aspect ratio on cross-flow turbine performance.
\newblock {\em Journal of Renewable and Sustainable Energy (in review)}.

\bibitem{hunt2023parametric}
Aidan Hunt, Benjamin Strom, Gregory Talpey, Hannah Ross, Isabel Scherl, Steven
  Brunton, Martin Wosnik, and Brian Polagye.
\newblock A parametric evaluation of the interplay between geometry and scale
  on cross-flow turbine performance.
\newblock {\em arXiv preprint arXiv:2310.20616}, 2023.

\bibitem{jin2020oe}
Guoqing Jin, Zhi Zong, Yichen Jiang, and Li~Zou.
\newblock Aerodynamic analysis of side-by-side placed twin vertical-axis wind
  turbines.
\newblock {\em Ocean Engineering}, 209:107296, 2020.

\bibitem{jodai2023wind}
Yoshifumi Jodai and Yutaka Hara.
\newblock Wind-tunnel experiments on the interactions among a pair/trio of
  closely spaced vertical-axis wind turbines.
\newblock {\em Energies}, 16(3):1088, 2023.

\bibitem{kinzel_2015}
Matthias Kinzel, Daniel~B Araya, and John~O Dabiri.
\newblock Turbulence in vertical axis wind turbine canopies.
\newblock {\em Physics of Fluids}, 27(11):115102, 2015.

\bibitem{kinzel2012jot}
Matthias Kinzel, Quinn Mulligan, and John~O Dabiri.
\newblock Energy exchange in an array of vertical-axis wind turbines.
\newblock {\em Journal of Turbulence}, 13(1):N38, 2012.

\bibitem{lam2017energy}
HF~Lam and HY~Peng.
\newblock Measurements of the wake characteristics of co-and counter-rotating
  twin h-rotor vertical axis wind turbines.
\newblock {\em Energy}, 131:13--26, 2017.

\bibitem{le2022dynamic}
S{\'e}bastien Le~Fouest and Karen Mulleners.
\newblock The dynamic stall dilemma for vertical-axis wind turbines.
\newblock {\em Renewable Energy}, 198:505--520, 2022.

\bibitem{miller2016dfd}
Michael Miller, Jennifer Cardona, Leanne Block, Kenta Kondo, Michael Lee,
  Rebecca Lorick, Michael Manning, Isabel Scherl, Filip Simeski, Arriane
  Spaulding, et~al.
\newblock Results from the field test of two 1 kw oscillating hydrofoil
  generators in a tidal canal.
\newblock In {\em APS Division of Fluid Dynamics Meeting Abstracts}, pages
  M2--009, 2016.

\bibitem{parker2016eif}
Colin~M Parker and Megan~C Leftwich.
\newblock The effect of tip speed ratio on a vertical axis wind turbine at high
  reynolds numbers.
\newblock {\em Experiments in Fluids}, 57(5):74, 2016.

\bibitem{polagye2019jrse}
Brian Polagye, Ben Strom, Hannah Ross, Dominic Forbush, and Robert~J Cavagnaro.
\newblock Comparison of cross-flow turbine performance under torque-regulated
  and speed-regulated control.
\newblock {\em Journal of Renewable and Sustainable Energy}, 11(4):044501,
  2019.

\bibitem{portugal2014nature}
Steven~J Portugal, Tatjana~Y Hubel, Johannes Fritz, Stefanie Heese, Daniela
  Trobe, Bernhard Voelkl, Stephen Hailes, Alan~M Wilson, and James~R Usherwood.
\newblock Upwash exploitation and downwash avoidance by flap phasing in ibis
  formation flight.
\newblock {\em Nature}, 505(7483):399--402, 2014.

\bibitem{posa2019ijhf}
Antonio Posa.
\newblock Wake characterization of coupled configurations of vertical axis wind
  turbines using large eddy simulation.
\newblock {\em International Journal of Heat and Fluid Flow}, 75:27--43, 2019.

\bibitem{posa2022re}
Antonio Posa.
\newblock Wake characterization of paired cross-flow turbines.
\newblock {\em Renewable Energy}, 2022.

\bibitem{posa2016johff}
Antonio Posa, Colin~M Parker, Megan~C Leftwich, and Elias Balaras.
\newblock Wake structure of a single vertical axis wind turbine.
\newblock {\em International Journal of Heat and Fluid Flow}, 61:75--84, 2016.

\bibitem{ribeiro2021prf}
Bernardo Luiz~R Ribeiro, Yunxing Su, Quentin Guillaumin, Kenneth~S Breuer, and
  Jennifer~A Franck.
\newblock Wake-foil interactions and energy harvesting efficiency in tandem
  oscillating foils.
\newblock {\em Physical Review Fluids}, 6(7):074703, 2021.

\bibitem{ross2020re}
Hannah Ross and Brian Polagye.
\newblock An experimental assessment of analytical blockage corrections for
  turbines.
\newblock {\em Renewable Energy}, 152:1328--1341, 2020.

\bibitem{ryan2016eif}
Kevin~J Ryan, Filippo Coletti, Christopher~J Elkins, John~O Dabiri, and John~K
  Eaton.
\newblock Three-dimensional flow field around and downstream of a subscale
  model rotating vertical axis wind turbine.
\newblock {\em Experiments in Fluids}, 57(3):38, 2016.

\bibitem{sahebzadeh2020towards}
Sadra Sahebzadeh, Abdolrahim Rezaeiha, and Hamid Montazeri.
\newblock Towards optimal layout design of vertical-axis wind-turbine farms:
  Double rotor arrangements.
\newblock {\em Energy Conversion and Management}, 226:113527, 2020.

\bibitem{salter2012coe}
SH~Salter.
\newblock Are nearly all tidal stream turbine designs wrong?
\newblock In {\em 4th International Conference on Ocean Energy}, pages 1--7,
  2012.

\bibitem{scherl2022optimization}
Isabel Scherl.
\newblock {\em Optimization, Modeling, and Control of Cross-Flow Turbine
  Arrays}.
\newblock PhD thesis, University of Washington, 2022.

\bibitem{scherlparameter}
Isabel Scherl, Steven~L Brunton, and Brian~L Polagye.
\newblock Parameter modeling of a two cross-flow turbine array from
  experimental data.

\bibitem{scherl2020jrse}
Isabel Scherl, Benjamin Strom, Steven~L Brunton, and Brian~L Polagye.
\newblock Geometric and control optimization of a two cross-flow turbine array.
\newblock {\em Journal of Renewable and Sustainable Energy}, 12(6):064501,
  2020.

\bibitem{simao2009visualization}
Carlos Sim{\~a}o~Ferreira, Gijs Van~Kuik, Gerard Van~Bussel, and Fulvio
  Scarano.
\newblock Visualization by piv of dynamic stall on a vertical axis wind
  turbine.
\newblock {\em Experiments in fluids}, 46:97--108, 2009.

\bibitem{snortland2023cycle}
Abigale Snortland, Isabel Scherl, Brian Polagye, and Owen Williams.
\newblock Cycle-to-cycle variations in cross-flow turbine performance and flow
  fields.
\newblock {\em arXiv preprint arXiv:2302.03218}, 2023.

\bibitem{strom2017natenergy}
B.~Strom, S.~L. Brunton, and B.~Polagye.
\newblock Intracycle angular velocity control of cross-flow turbines.
\newblock {\em Nature Energy}, 2(17103):1--9, 2017.

\bibitem{strom2018jrse}
Benjamin Strom, Noah Johnson, and Brian Polagye.
\newblock Impact of blade mounting structures on cross-flow turbine
  performance.
\newblock 2018.

\bibitem{strom2022jfm}
Benjamin Strom, Brian Polagye, and Steven~L Brunton.
\newblock Near-wake dynamics of a vertical-axis turbine.
\newblock {\em Journal of Fluid Mechanics}, 935, 2022.

\bibitem{tescione2014re}
G~Tescione, D~Ragni, C~He, CJ~Sim{\~a}o Ferreira, and GJW Van~Bussel.
\newblock Near wake flow analysis of a vertical axis wind turbine by
  stereoscopic particle image velocimetry.
\newblock {\em Renewable Energy}, 70:47--61, 2014.

\bibitem{vergaerde2020re}
Antoine Vergaerde, Tim De~Troyer, Lieven Standaert, Joanna Kluczewska-Bordier,
  Denis Pitance, Alexandre Immas, Fr{\'e}d{\'e}ric Silvert, and Mark~C
  Runacres.
\newblock Experimental validation of the power enhancement of a pair of
  vertical-axis wind turbines.
\newblock {\em Renewable Energy}, 146:181--187, 2020.

\bibitem{verma2018pnas}
Siddhartha Verma, Guido Novati, and Petros Koumoutsakos.
\newblock Efficient collective swimming by harnessing vortices through deep
  reinforcement learning.
\newblock {\em Proceedings of the National Academy of Sciences},
  115(23):5849--5854, 2018.

\bibitem{weis1973quick}
Torkel Weis-Fogh.
\newblock Quick estimates of flight fitness in hovering animals, including
  novel mechanisms for lift production.
\newblock {\em Journal of experimental Biology}, 59(1):169--230, 1973.

\bibitem{whelan2009free}
Jo~I Whelan, JMR Graham, and Joaquim Peiro.
\newblock A free-surface and blockage correction for tidal turbines.
\newblock {\em Journal of Fluid Mechanics}, 624:281--291, 2009.

\bibitem{whittlesey2010bioinspiration}
Robert~W Whittlesey, Sebastian Liska, and John~O Dabiri.
\newblock Fish schooling as a basis for vertical axis wind turbine farm design.
\newblock {\em Bioinspiration \& biomimetics}, 5(3):035005, 2010.

\bibitem{wu2011arfm}
Theodore~Yaotsu Wu.
\newblock Fish swimming and bird/insect flight.
\newblock {\em Annual Review of Fluid Mechanics}, 43:25--58, 2011.

\bibitem{zanforlin2016re}
Stefania Zanforlin and Takafumi Nishino.
\newblock Fluid dynamic mechanisms of enhanced power generation by closely
  spaced vertical axis wind turbines.
\newblock {\em Renewable Energy}, 99:1213--1226, 2016.

\end{thebibliography}
\bibliographystyle{plain}

\end{document}